# Oxide-Nitride Heteroepitaxy for Low-Loss Dielectrics in Superconducting Quantum Circuits

*David A. Garcia-Wetten[1,†,‡], Mitchell J. Walker[1,†,‡], Peter G. Lim[2,‡], André Vallières[2,3,‡], Maria G. Jimenez-Guillermo[4], Miguel A. Alvarado[5], Dominic P. Goronzy[1], Anna Grassellino[3], Jens Koch[6,7], Vinayak P. Dravid[1,8,9,*], Mark C. Hersam[1,10,11,*], and Michael J. Bedzyk[1,6,*]*

[1]Department of Materials Science and Engineering, Northwestern University, Evanston, IL 60208, USA

[2]Graduate Program in Applied Physics, Northwestern University, Evanston, IL 60208, USA

[3]Superconducting Quantum Materials and Systems (SQMS) Division, Fermi National Accelerator Laboratory (FNAL), Batavia, IL 60510, USA

[4]Department of Physics, Elmhurst University, Elmhurst, IL 60126, USA

[5]Department of Chemistry, Northeastern Illinois University, Chicago, IL 60625, USA

[6]Department of Physics and Astronomy, Northwestern University, Evanston, IL 60208, USA

[7]Center for Applied Physics and Superconducting Technologies (CAPST), Northwestern University, Evanston, IL 60208, USA

[8]Northwestern University Atomic and Nanoscale Characterization Experimental (NU*ANCE*) Center, Northwestern University, Evanston, IL 60208, USA

[9]International Institute of Nanotechnology (IIN), Northwestern University, Evanston, IL 60208, USA

[10]Department of Chemistry, Northwestern University, Evanston, IL 60208, USA

[11]Department of Electrical and Computer Engineering, Northwestern University, Evanston, IL 60208, USA

**AUTHOR INFORMATION**

*Corresponding authors:

v-dravid@northwestern.edu, m-hersam@northwestern.edu, bedzyk@northwestern.edu

‡These authors contributed equally to this work.

**ABSTRACT**

Superconducting qubits show great promise for the realization of fault-tolerant quantum computing, but lossy, amorphous dielectrics limit current technology. Identifying highly crystalline and stoichiometric dielectrics with intrinsically low microwave loss is therefore a central materials challenge, yet experimentally validated platforms remain scarce. This work integrates a crystalline dielectric into a heteroepitaxial TiN/γ-$Al_2O_3$/TiN trilayer grown *via* pulsed laser deposition. Correlative high-resolution imaging, diffraction, and spectroscopy measurements confirm the single-crystal quality and chemical integrity of all layers, with minimal defects and limited anion interdiffusion across the oxide-nitride interfaces. Using microwave lumped-element resonators with parallel-plate capacitors, this work establishes an experimentally-measured upper bound for the dielectric loss of epitaxial γ-$Al_2O_3$, which has a low intrinsic two-level system loss of $\delta^0_{\mathrm{TLS}} = (2.8 \pm 0.1) \times 10^{-5}$. These results establish heteroepitaxial oxides on transition metal nitrides as an attractive materials platform for superconducting quantum circuits, particularly for integration into compact device architectures such as merged-element transmons and microwave kinetic inductance detectors.

Since the proposition[1] and subsequent demonstration[2, 3] of Cooper-pair box qubits, superconducting (SC) qubit technology has accelerated with broad potential for quantum information science (QIS) applications. The incumbent materials choice for the Josephson junctions (JJs) in these devices is amorphous $AlO_x$ for the dielectric and Al for the SC electrodes. However, amorphous $AlO_x$ as the tunnel barrier is problematic since it hosts parasitic two-level systems (TLSs),[4, 5] which are a major source of decoherence in SC qubits.[6-8] The lack of long-range order and sub-stoichiometry of amorphous oxides give rise to a wide distribution of atomic coordination geometries,[9] which implies that these defective materials host TLSs with energetically similar configurations to the SC qubits themselves, ultimately compromising coherence.[10] Because single-crystal dielectrics lack the disorder of amorphous materials and thus have a lower density of TLSs, layer-by-layer growth of atomically pristine dielectrics *via* thin-film epitaxy is an attractive alternative to incumbent amorphous $AlO_x$.

Several material systems have been studied for this purpose. For example, epitaxial semiconductors, such as GaAs,[11, 12] have been used with some success to reduce TLS losses. Another strategy is to fabricate an all-nitride trilayer stack using AlN and transition-metal nitride superconductors.[13, 14] However, both AlN and GaAs are susceptible to higher power-independent losses due to their piezoelectricity.[15, 16] An alternative approach is to form a crystalline $Al_2O_3$ dielectric by depositing at high temperatures. In this strategy, the choice of the SC base layer is key to limiting metal oxidation *via* interdiffusion with $Al_2O_3$ at elevated growth temperatures. For instance, oxygen interdiffusion at the Nb/$Al_2O_3$ interface precludes the use of Nb for this purpose.[17] Therefore, prior attempts at epitaxial growth of crystalline $Al_2O_3$ have incorporated Re[18] or Re/Ti multilayers,[19, 20] but Re is a suboptimal superconductor for qubit devices due to its poor coherence, low SC energy gap, and high kinetic inductance.[21] Instead, transition metal nitrides stand out as promising superconductors for QIS applications in light of these criteria.[22-26] Towards this end, the growth of $Al_2O_3$/NbN trilayers has been investigated, but this approach is limited by oxygen and nitrogen interdiffusion.[27] TiN is particularly attractive in this context because it is a low-loss superconductor that is already widely integrated into QIS technologies[28-34] and related applications, owing to its resistance to oxidation and effectiveness as a diffusion barrier.[35, 36]

SC qubit technology additionally faces a scale-up problem. Conventional transmon qubits often incorporate large external shunt capacitors to mitigate electric field participation of lossy materials in the circuit.[37-39] Together with the readout elements, which are typically coplanar waveguide (CPW) resonators,[40] these devices can occupy areas on the order of 2 – 3 $mm^2$ per device, which limits integration density for many-qubit quantum processing units. Consequently, SC qubits with smaller footprints are of high interest, but existing platforms (*e.g.*, merged-element transmons[41, 42]) have low coherence times that are limited by constituent dielectrics. Prior attempts to address this problem, such as using van der Waals material[43, 44] as the dielectric, while dramatically reducing the device footprint, still rely on techniques incompatible with industrial-scale device fabrication. Consequently, low-loss epitaxial dielectrics grown with wafer-scale compatible thin-film epitaxy are essential not only for improving SC qubit coherence but also for the development of many-qubit quantum processing units.

Here, we demonstrate the growth of high-quality epitaxial TiN (111)/$\gamma$-$Al_2O_3$ (111)/TiN (111) trilayers on $\alpha$-$Al_2O_3$ (001) substrates *via* pulsed laser deposition (PLD). PLD is chosen because it is a deposition method that has previously yielded high-quality crystalline films of both TiN[45] and $Al_2O_3$.[46] In this manner, all three films of the trilayer can be grown in the same deposition system without breaking vacuum or requiring a multi-use system. X-ray reflectivity (XRR), X-ray diffraction (XRD), time-of-flight secondary ion mass spectrometry (ToF-SIMS), and scanning transmission electron microscopy (STEM) are used to characterize layer thicknesses, interfaces, and crystallinities. X-ray photoelectron spectroscopy (XPS) and electron energy-loss spectroscopy (EELS) are further employed to characterize the chemical state of the constituent films. Through these robust structural and spectroscopic measurements, we find that each layer of the stack is highly stoichiometric and crystalline, while collectively maintaining heteroepitaxy, with sharp interfaces and minimal oxygen-nitrogen interdiffusion. To accurately measure the TLS contribution to the intrinsic material loss of epitaxial γ-$Al_2O_3$, we employ a lumped-element resonator,[47] which minimizes the capacitance of the inductor element to ground. By integrating this with a parallel-plate capacitor (PPC) with a filling factor close to unity, TLSs are strongly capacitively coupled in this design, allowing us to attribute the measured TLS losses predominantly to the capacitor dielectric. We test devices with varying dielectric thicknesses to test whether the measurements reflect the bulk γ-$Al_2O_3$'s loss properties or those of the TiN/γ-$Al_2O_3$ interfaces. In this manner, we find that the TLSs loss of crystalline γ-$Al_2O_3$ in the epitaxial trilayer is as low as $\delta^0_{\mathrm{TLS}} = (2.8 \pm 0.1) \times 10^{-5}$, which is two orders of magnitude lower than conventional amorphous $AlO_x$. Our lumped-element resonators with an approximate area of $2.4 \times 10^{-2}$ $mm^2$ — a reduction by two orders of magnitude compared to CPW geometries — exhibit exceptionally high quality factors that hold significant promise for emerging QIS applications.

## RESULTS AND DISCUSSION

### Growth and Characterization of TiN/γ-$Al_2O_3$/TiN Heteroepitaxial Trilayers

Two trilayers of TiN/$Al_2O_3$/TiN are deposited onto c-plane sapphire substrates by PLD. From top to bottom, the first sample (Figure 1a) is 62.2 nm TiN - 13.5 nm $Al_2O_3$ - 53.8 nm TiN, and the second sample (Figure 1c) is 63.0 nm TiN - 58.3 nm $Al_2O_3$ - 68.1 nm TiN, as measured by STEM with ± 0.5 nm uncertainty. Energy dispersive X-ray spectroscopy (EDS) mapping (Figure 1b and 1d) confirms that for both the thin and the thick $Al_2O_3$ samples, the elemental distribution is qualitatively as expected, with no significant signs of oxygen in the TiN layers, and that the interfaces are sharp and well-defined.

Epitaxy during PLD growth is verified with reflectance high-energy electron diffraction (RHEED). The RHEED pattern (Figure 1e) shows vertical streaks with faint spots, indicating some island formation, but predominantly sharp interfaces,[48] consistent with the STEM findings. Layer thicknesses extracted from XRR measurements (Figure 1f) over a larger area of the two trilayer films is also consistent with STEM measurements: the measured $Al_2O_3$ layer thicknesses are 13.5 nm for the first sample (green curve in Figure 1f) and 58.3 nm for the second sample (blue curve in Figure 1f). Additionally, the scattering length densities of the TiN and $Al_2O_3$ (inset of Figure 1f) are consistent with their respective theoretical values. Depth-profiling ToF-SIMS of the thick $Al_2O_3$ sample (Figure 1g) also reveals sharp interfaces at both TiN/$Al_2O_3$ interfaces and minimal interdiffusion between the layers. The peak in the $TiNO^-$ signal at 0 s of etch time is consistent with a thin layer of oxidized TiN at the top of the trilayer stack, which is exposed to air.

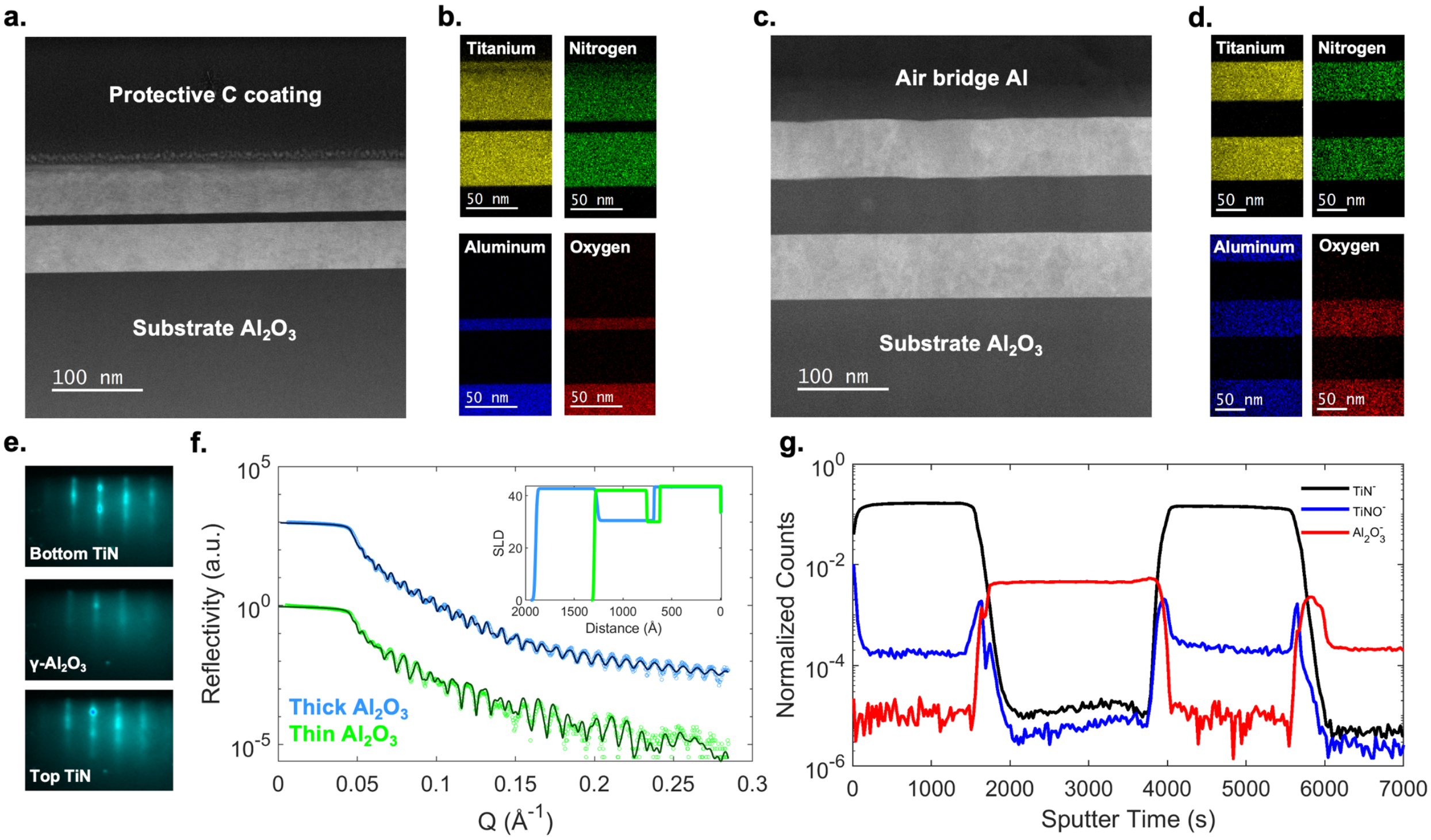

**Figure 1.** Materials characterization of PLD-deposited $TiN/Al_2O_3/TiN$ trilayers. (a,c) HAADF-STEM images and (b,d) EDS maps of the thin and thick trilayer samples, showing well-defined, elementally homogeneous layers with flat interfaces. (e) *in situ* RHEED of the surface of the bottom TiN, middle $Al_2O_3$, and top TiN layers, showing epitaxial matching with some island growth. (f) XRR of the thin and thick trilayer samples. The inset shows the scattering-length density profile from the fits. (g) ToF-SIMS depth profile of the thick trilayer sample, showing the depth-distribution of $TiN^-$, $TiON^-$, and $Al_2O_3^-$ species.

To investigate the crystal structures and epitaxial relationships of all layers, we perform high-resolution STEM and high-resolution XRD. Annular bright-field (ABF-)STEM images (Figure 2a, 2b, and 2c) of the trilayers additionally demonstrate that both TiN layers have the expected rock salt structure in the (111) specular orientation (insets in Figure 2a and 2c), and in particular, that the bottom TiN layer grows epitaxially to the c-plane sapphire substrate. Both $TiN/Al_2O_3$ interfaces are also flat and epitaxial, albeit with some roughness attributed to anion interdiffusion, as discussed later. Of note, the alumina (Figure 2b) is a single-crystal with the cubic $\gamma$-$Al_2O_3$ phase, in contrast to the hexagonal-R $\alpha$-$Al_2O_3$ sapphire substrate, as evidenced by the appearance of superlattice peaks in the fast Fourier transform (FFT) patterns (inset in Figure 2b) and four-dimensional (4D-)STEM diffraction patterns (Figure S1). The γ-$Al_2O_3$ phase is a defective spinel-like cubic $Fd\bar{3}m$ crystal with both tetrahedral and octahedral coordination of the Al sites, enabled by ordered oxygen vacancies,[49] which grows epitaxially on cubic $Fm\bar{3}m$ TiN crystals at elevated temperatures (Figure 2d).

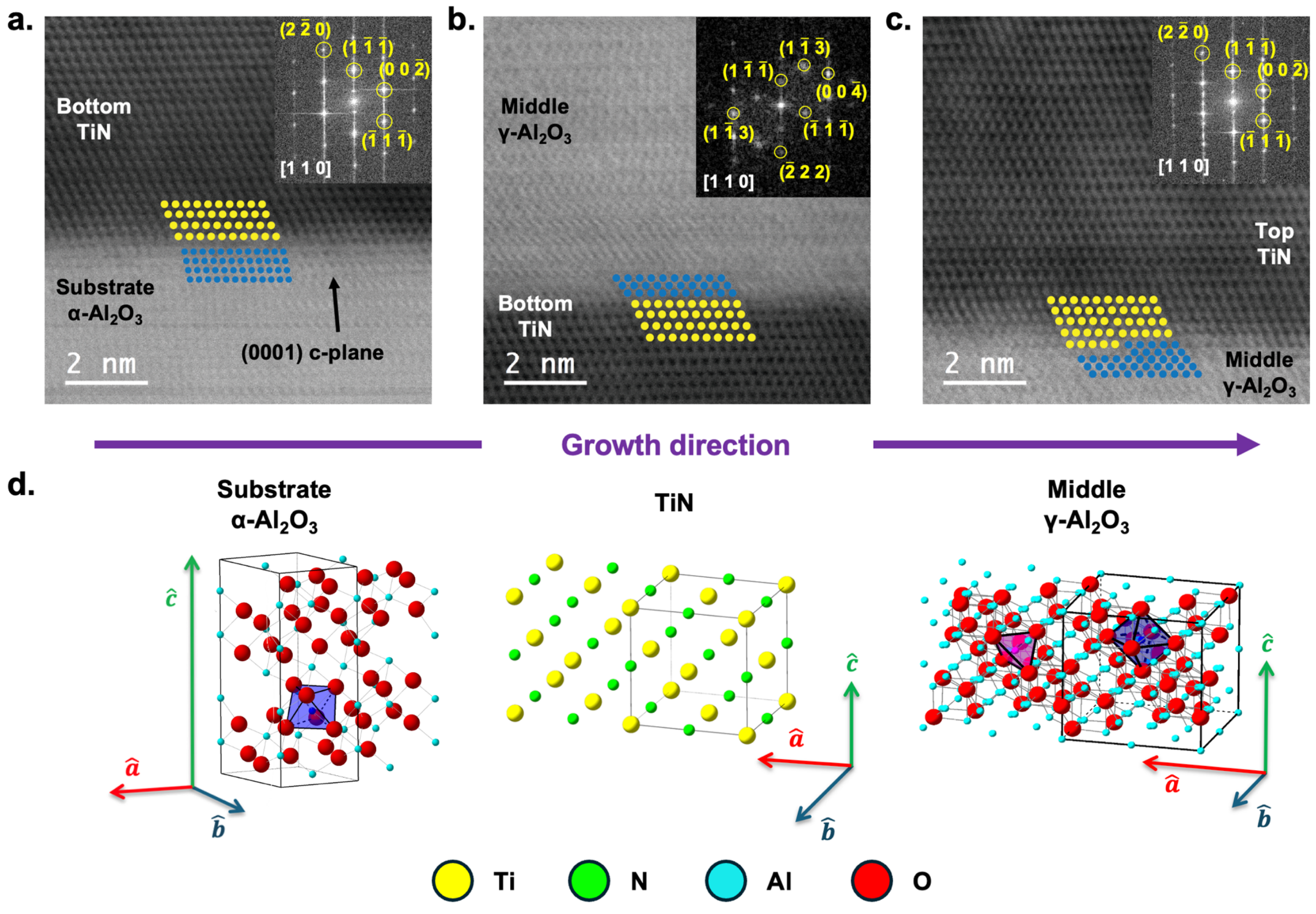

**Figure 2.** Crystal structures and interfaces of TiN/γ-$Al_2O_3$/TiN. (a-c) Atomic-resolution ABF-STEM images of the interfaces between each film layer and the substrate, demonstrating heteroepitaxy and high-quality interfaces. The insets show FFT patterns of the individual layers, with the zone axes and diffraction peaks identified and labeled. (d) Simulated crystal structures of the substrate and films.

XRD provides further information on the heteroepitaxial relationship of the TiN and γ-$Al_2O_3$ layers. First, referring to the specular scattering pattern (purple curve in Figure 3a), the TiN (111) and (222) appear at $Q = 2.56$ Å$^{-1}$ and 5.10 Å$^{-1}$, respectively. The γ-$Al_2O_3$ (222) and (444) are present at $Q = 2.75$ Å$^{-1}$ and 5.48 Å$^{-1}$, but the (333) is missing. Unlike the (222) reflection, which mainly originates from the well-ordered oxygen sublattice, the (333) reflection is typically much broader due to disorder on the Al sites.[50] This omission is not unprecedented in epitaxial γ-$Al_2O_3$ thin films and may arise from a difference in cation ordering from the bulk crystal structure.[51] This discrepancy suggests that the growth method plays a role in the amount of disorder on the Al sites within the bulk of the γ-$Al_2O_3$ and may contribute to bulk dielectric loss. The peak at $Q$ = 3.08 $Å^{-1}$ is TiN (002). This "impurity" orientation is present only in the top TiN layer, as shown by virtual dark-field images in the 4D-STEM measurements (Figure S1f and S1g).

Off-specular, radial Q scans (green curves in Figure 3a) confirm the cubic structure of γ-$Al_2O_3$. At an elevation of 54.7º in the cubic [110] direction, the TiN (220) appears at $Q$ = 2.41 $Å^{-1}$, as well as γ-$Al_2O_3$ (220) and (440) at $Q$ = 2.25 $Å^{-1}$ and 4.50 $Å^{-1}$, respectively. At an elevation of 35.3º in the cubic [001] direction, the TiN (002) is visible at $Q$ = 2.96 $Å^{-1}$, and the γ-$Al_2O_3$ (004) is visible at 3.18 $Å^{-1}$. Additionally, the γ-$Al_2O_3$ (113) is observable at an elevation of 60.5º and $Q$ = 2.62 $Å^{-1}$. This peak is a superlattice reflection from the superstructure of ordered Al tetrahedral and octahedral sites in the γ-$Al_2O_3$ structure. All of these peaks are observed to have 6-fold symmetry about the specular direction, indicating a 180º-twin (Figure 3b). This twinning is expected for cubic (111)/sapphire (001) epitaxy, which may arise because of alternating A-B stacking from step-to-step on the sapphire surface.[52]

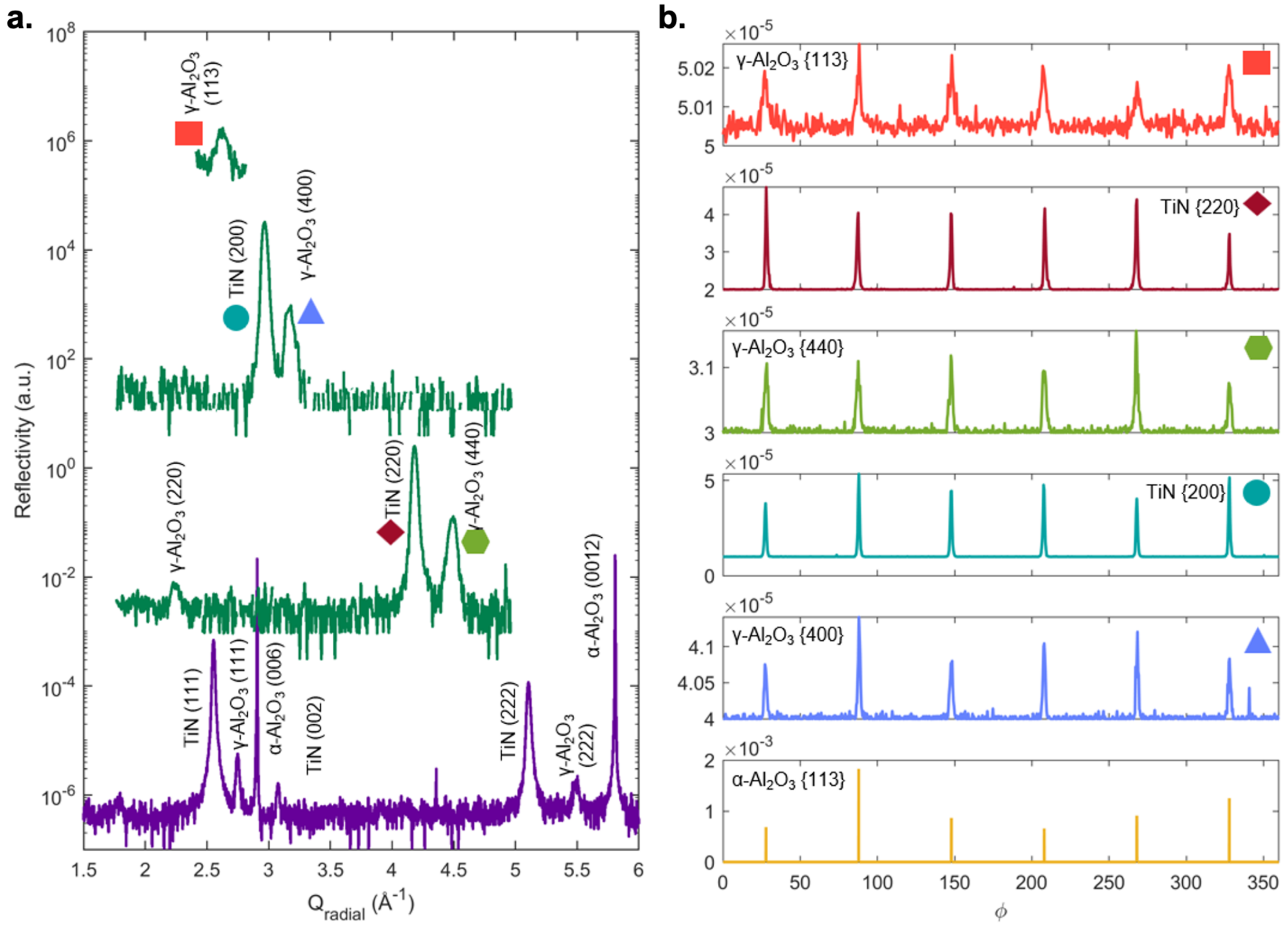

**Figure 3.** Heteroepitaxy and crystal symmetries of TiN/γ-$Al_2O_3$/TiN. (a) Specular (purple) and off-specular, radial Q (green) XRD scans of the thick trilayer sample, and (b) $\varphi$ scans of selected XRD peaks, showing the rotational symmetry and epitaxial matching of the film layers.

Chemical-state characterization of the trilayer stacks is performed using XPS and EELS. The surface XPS scans (dark blue curves in Figure 4a) show that the top surface is composed of TiN, reflected by the strong first doublet in the Ti 2p spectrum (the "TiN doublet") and the tallest peak at 395 eV in the N 1s spectrum.[53] The TiN doublet is separated from the N 1s signal by 57.7 eV, indicative of TiN with some N deficiency.[54] This observation is consistent with the prominent $TiO_x$ peak shifted by 3.1 eV from the TiN doublet and the strong O 1s signal;[55] these features are attributable to oxidation of the TiN surface from ambient exposure, in agreement with the ToF-SIMS data (Figure 1g). The remaining peaks involved in the Ti 2p fitting can be interpreted as two $Ti_xO_yN_z$ doublets (based on their separations of 1.1 eV and 2.3 eV from the TiN doublet),[55] shake-up satellite features, surface plasmon loss features, and bulk plasmon loss features. The shake-up peaks are fixed at 1.9x the area of the TiN peaks, with splitting according to Jaeger *et al.*[53] Two additional peaks are added to fit the shoulders in the N 1s spectrum: the 393.7 eV peak is associated with TiON,[55] while the 396.1 eV peak could not be assigned.

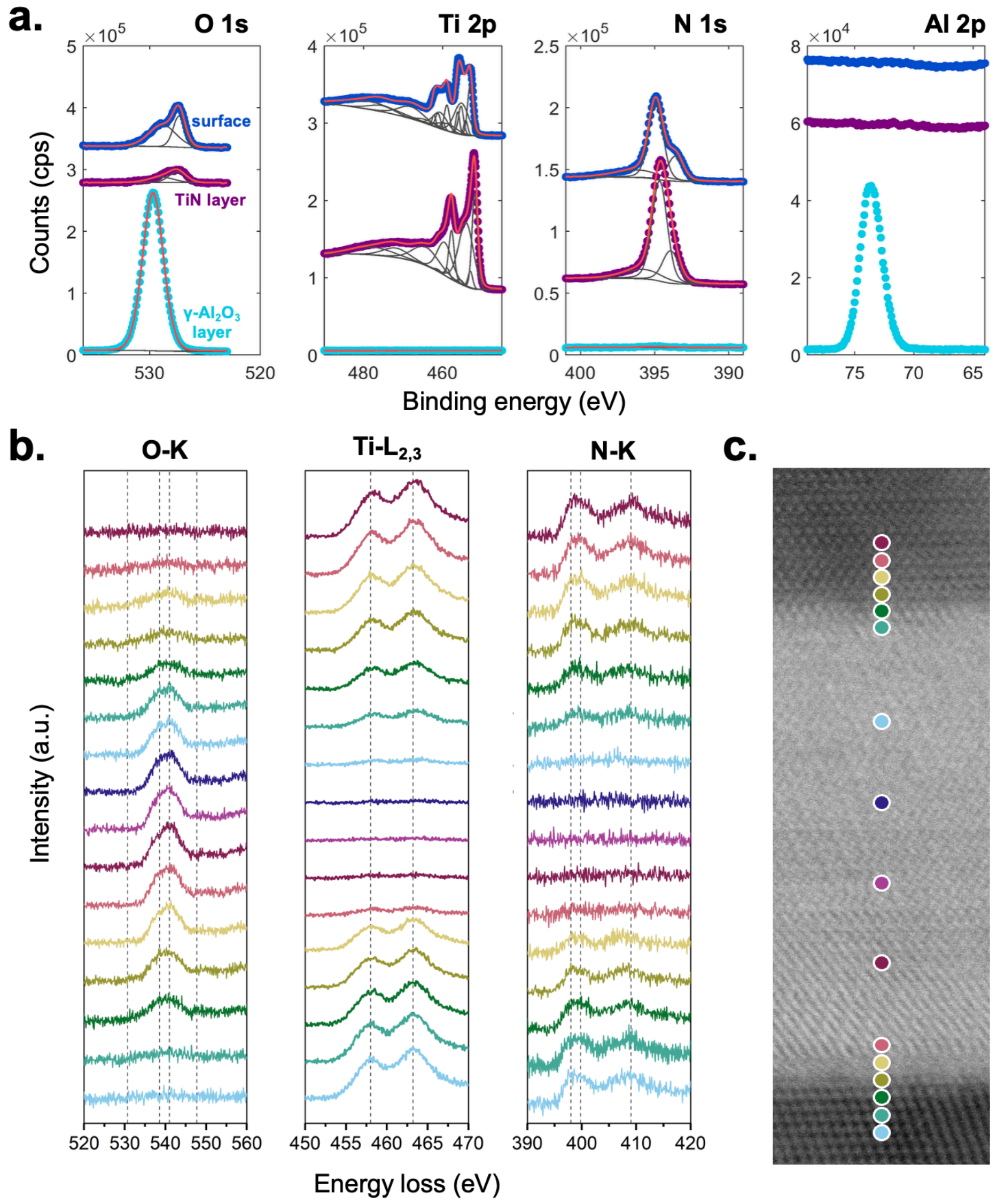

**Figure 4.** Chemical analysis at the TiN/γ-$Al_2O_3$ interfaces. (a) XPS of core-level emissions of the thick trilayer sample. Buried layers were probed using an $Ar^+$ ion beam. (b) Core-loss EELS along the two interfaces between the TiN layers and γ-$Al_2O_3$ film, with (c) an ABF-STEM image indicating the location of these spectra.

XPS can also probe the bulk and buried interfaces by sputtering the surface with $Ar^+$ ions between successive scans. The magenta curves in Figure 4a show XPS results taken at a depth midway through the top TiN layer. Referring to the Ti 2p and N 1s spectra, the ratio of $Ti_xO_yN_z$ and $TiO_x$ signal intensities to those of TiN is much weaker compared to the same ratios in the dark blue, surface-level XPS curves. Likewise, the O 1s signal coming from the bulk TiN is much weaker than the dark blue, surface-level O 1s signal. These findings are corroborated with elemental quantification, which indicates that the TiN bulk is close to stoichiometric with just 8 at.% O. This oxygen concentration is slightly higher than we would expect based on our SC transition temperature ($T_c$) measurement of 4.8 K (Figure S2), so we hypothesize that this percentage is an over-estimate due to $Ar^+$ ion milling damage and oxygen gettering in the high (but not ultrahigh) vacuum XPS chamber.[56]

The cyan curves in Figure 4a show XPS from a depth in the center of the $\gamma$-$Al_2O_3$ layer. Here, the Al 2p and O 1s signals dominate, as expected, with the Ti 2p and N 1s signals having disappeared almost entirely. No shoulders from metal oxides are present in the O 1s spectrum. Elemental quantification confirms that the $\gamma$-$Al_2O_3$ is highly stoichiometric.

Further interrogation of the chemical state of the $\gamma$-$Al_2O_3$ and its interfaces with the TiN is performed using EELS. The EELS data (Figure 4b) correspond to the O-K edge, Ti-$L_{2,3}$ edge, and N-K edge measured along the vertical direction of the dielectric and are color-matched with the ABF-STEM image (Figure 4c). The Ti-$L_{2,3}$ edge shape throughout the TiN in proximity to the $\gamma$-$Al_2O_3$ is predominantly of $Ti^{3+}$, corresponding to highly stoichiometric TiN. There are no signs of a significant Ti-$L_{2,3}$ edge shift or splitting indicative of $Ti^{4+}$ nor an O-K edge field splitting into the $t_{2g}$ and $e_g$ levels, which would indicate formation of $TiO_2$.[57] These findings confirm that the TiN layer is chemically resistant to oxidation, unlike many other standard SC electrodes used in qubits. Furthermore, the O-K edge spectrum throughout the bulk of the dielectric layer is in good agreement with the known spectrum of $\gamma$-$Al_2O_3$, with an edge shape clearly distinct from the O-K spectrum of $\alpha$-$Al_2O_3$,[58, 59] providing further corroboration of the identification of the $Al_2O_3$ as gamma phase.

Despite largely clean and flat interfaces, the splitting of the O-K edge into two peaks at ~541 eV and ~539 eV at both interfaces with the TiN resembles electronic transitions into mixed states of the O 2p and Al $t_{2g}$ and $e_g$ states, corresponding to both tetrahedral and octahedral coordination of the Al sites, suggesting increasingly disordered oxygen vacancy sites near the interfaces. An additional extended energy-loss fine structure at ~549 eV arises from transitions into mixed states of the O 2p and the Al 3s and 3p state.[55, 58] These measurements additionally reveal the formation of ~1.5 nm thin $Ti_xO_yN_z$ nonstoichiometric interlayers at both TiN/$\gamma$-$Al_2O_3$ interfaces, as observed by the coexistence of N-K, Ti-$L_{2,3}$, and O-K edges over a non-negligible region along the interfaces. Changes in the O vs. N content across the intermixing region is gradual, making it difficult to draw quantitative conclusions about the relative atomic concentration of the anions. Nevertheless, these $Ti_xO_yN_z$ interlayers can be attributed to oxygen migration from $\gamma$-$Al_2O_3$, evident from the appearance of an O-K pre-edge at ~532 eV indicating oxygen deficiency in $\gamma$-$Al_2O_3$, and the appearance of a secondary N-K edge at ~400 eV representing displaced nitrogen atoms dissolved as unbonded nitrogen in the matrix.[60] In other words, the interlayer is predominantly a $Ti_xN_z$ matrix with $O_y$ species migrating over from the $\gamma$-$Al_2O_3$. It is possible that some diffusion occurred at elevated temperatures during PLD deposition, which may explain the faint diffuseness visible at both interfaces in ABF-STEM images (Figure 2b and 2c).

**Dielectric Loss Extraction of Epitaxial γ-$Al_2O_3$**

To obtain an accurate measurement of the dielectric loss of epitaxial γ-$Al_2O_3$, lumped-element resonators with parallel-plate capacitors (LEPPCs) are designed, simulated, and fabricated. This device design concentrates the electric field in the PPCs, such that the majority of dielectric losses observed are dominated by either the bulk of the dielectric or its interfaces with the TiN layers. As depicted in Figure 5a and Figure S3, starting from a blanket trilayer film, the top TiN and the γ-$Al_2O_3$ are first patterned and etched into a circular-shaped capacitor. A second patterning and etching step defines the inductor, ground plane, and the feedline in the bottom TiN layer. An Al airbridge is used to connect the top electrode of the capacitor to the inductor, rather than an insulating spacer, to limit dielectric losses outside the capacitor dielectric.[12, 61] The resulting devices, as shown in scanning electron microscopy (SEM) images (Figure 5b and 5c), exhibit high fidelity to the desired device geometry, with no observable photoresist or other residue. The slight waviness on either side of the airbridge pictured in Figure 5b is likely due to the laser of the direct optical writer having imperfect focus over variations in sample height during photolithography. The interface between the trilayer and the Al airbridge is also sharp and free of oxygen and carbon residues (Figure S4). Figure 5d shows a simulation of the electric field distribution in the LEPPC device, derived from finite element modeling (FEM). The FEM model shows that the field is strongly concentrated in the PPC, with a filling factor near unity. The surface participation ratio around the perimeter of the PPC is negligible ($< 10^{-9}$) due to the large dimension ratio between the dielectric diameter and thickness. Therefore, the measured TLS loss of the entire device can be attributed to the intrinsic material TLS loss of the capacitor dielectric, assuming TLSs are only capacitively coupled.[47] Additional details of the device design, simulation, fabrication, and measurement are provided in the Methods and Supporting Information.

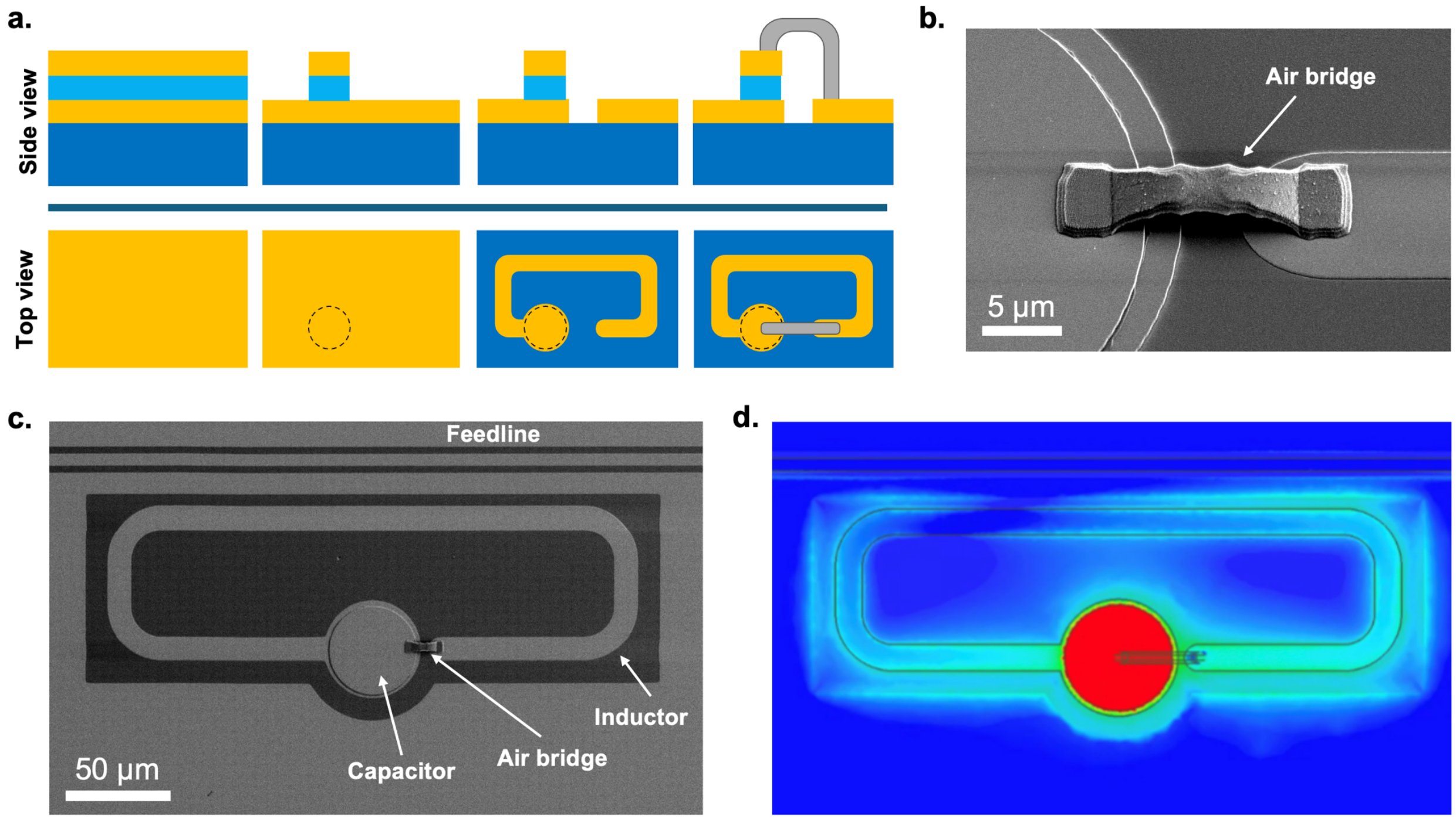

**Figure 5.** LEPPC device overview. (a) Schematic of LEPPC device fabrication procedure, where the top TiN (yellow) and dielectric γ-$Al_2O_3$ (light blue) are optically defined to be the PPC, and the bottom TiN (yellow) comprises the LE resonator, along with an aluminum airbridge (grey) completing the LC circuit. (b) Tilted SEM image of a representative airbridge, and (c) top-down SEM image of a representative LEPPC device. (d) FE modeling of the same device, displaying the electric field distribution upon microwave measurements concentrated in the PPC region. The color scale corresponds to the strength of the electric field on a logarithmic scale with arbitrary units (red - high, blue - low).

After fabrication, the devices are loaded into a dilution refrigerator at approximately 10 mK for microwave testing. The measured data points and TLS model fits[62] for both the LEPPC devices and co-fabricated CPW resonators are shown in Figure 6. At low powers, devices fabricated with the thin (13.5 nm) and the thick (58.3 nm) γ-$Al_2O_3$ perform similarly: the thick γ-$Al_2O_3$ device has a low-power total loss $\delta_{LP}$ of $(3.6 \pm 0.3) \times 10^{-5}$, while the thin γ-$Al_2O_3$ device has a $\delta_{LP}$ of $(3.2 \pm 0.2) \times 10^{-5}$, corresponding to the plateau from the model fits in the TLS-dominated regime at $\langle n \rangle \sim 10^{-4}$. While the relative electric field density at the TiN/γ-$Al_2O_3$ interfaces is much higher in the thin γ-$Al_2O_3$ device, we find that measured $\delta_{LP}$ is similar for both thicknesses. This suggests that the low-power loss is dominated by the bulk γ-$Al_2O_3$ rather than lossy elements at the TiN/γ-$Al_2O_3$ interfaces, such as the thin $Ti_xO_yN_z$ interlayers. Furthermore, the volume of the dielectric material being tested in each device is sufficiently large, such that the number of TLSs present is expected to match the theoretical TLS density for this material in the bulk limit. Therefore, these $\delta_{LP}$ values represent an upper bound—given the limited number of measured samples—of the low-power material loss of the γ-$Al_2O_3$ itself. By subtracting the fitted high-power loss plateau $\delta_{HP}$ from $\delta_{LP}$, we also extract $F\delta^0_{\mathrm{TLS}}$ of both devices. Based on prior discussions of our device design, $F$ is very close to 1, so these numbers directly represent the intrinsic TLS loss, $\delta^0_{\mathrm{TLS}}$, for γ-$Al_2O_3$ of both thicknesses, which we measure to be (3.5 ± 0.2) × $10^{-5}$ and (2.8 ± 0.1) × $10^{-5}$ for the thick and thin γ-$Al_2O_3$, respectively. Assuming the thin $Ti_xO_yN_z$ interlayers species are metallic and do not contribute to dielectric loss, these TLS loss values reflect the intrinsic dielectric loss of our γ-$Al_2O_3$ material.

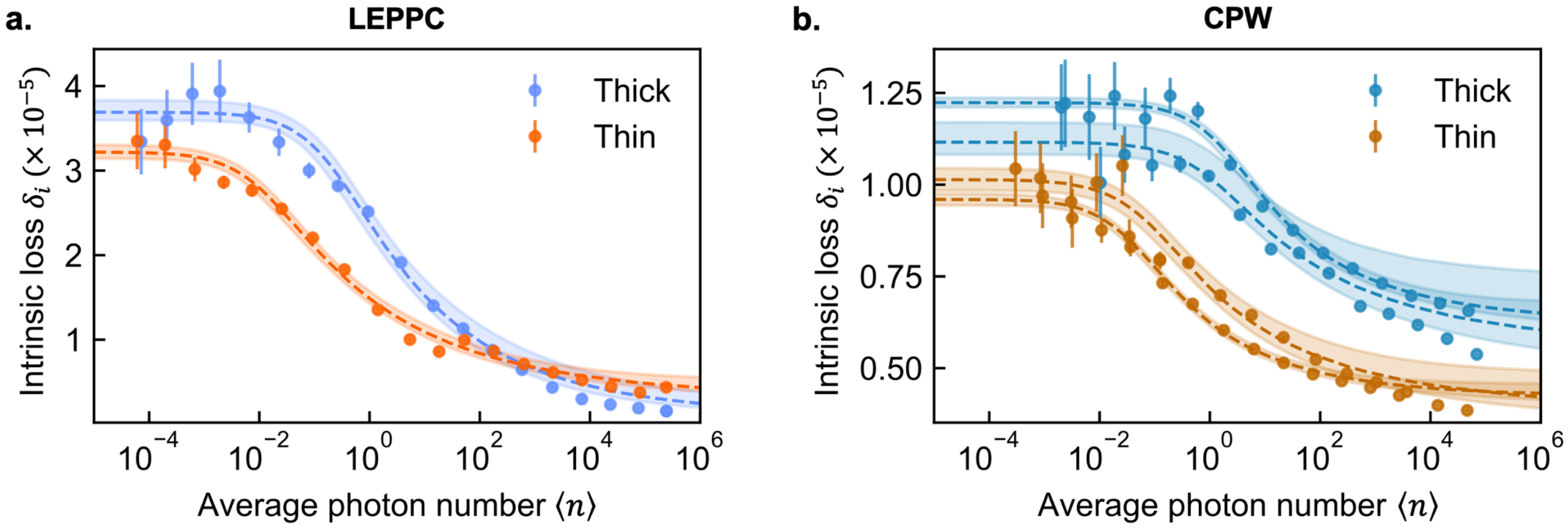

**Figure 6.** Cryogenic microwave measurement results of the resonators. Power-dependent intrinsic loss $\delta_i = 1/Q_i$ for (a) LEPPC and (b) CPW resonators patterned on the bottom TiN layer of the thick (blue) and thin (orange) dielectric devices. Shaded bands represent uncertainty derived from the covariance matrix of each fit.

At high powers, we measure exceptionally high internal quality factors, with maximum $Q_{max} = (6.4 \pm 0.5) \times 10^5$ for the thick and $Q_{max} = (2.7 \pm 0.1) \times 10^5$ for the thin γ-$Al_2O_3$ device, before the resonator response becomes non-linear. The significant difference in performance at high power suggests a power-independent loss mechanism dominated by the interfaces or elsewhere in the PPC[33]. This loss could be associated with the thin nonstoichiometric $Ti_xO_yN_z$ layers at the TiN/γ-$Al_2O_3$ interfaces observable by EELS (Figure 4b) and ToF-SIMS (Figure 1g). Prior work from the literature establish that $Ti_xO_yN_z$ is primarily conductive but becomes semiconducting as oxygen content increases.[63-66] Analysis of the EELS data is in agreement with $Ti_xO_yN_z$ species at the interfaces being nitrogen-rich, which is why they likely contribute to power-independent loss as conductor loss instead of dielectric TLS loss. Another possibility is that the loss could be associated with the twinned TiN domains revealed by 4D-STEM measurements (Figure S1f and S1g). These inhomogeneities in the SC film have been linked to degraded microwave performance by acting as preferential pathways for oxidation and impurity diffusion through the bulk metal, potentially inducing conductive losses.[67-69]

Table 1 compares these performance metrics with those of other dielectrics reported in the literature. Inclusion in this table is limited to measurements using integrated PPC-type geometries to keep comparisons meaningful; otherwise, filling factors and extraction protocols may differ widely. In general, $AlO_x$ can exhibit a wide range of loss values, depending on deposition technique, crystallinity, stoichiometry, and potentially other factors. Since most state-of-the-art JJs consist of amorphous $AlO_x$ formed by oxidizing electron-beam-deposited Al, among the reports listed in Table 1, Zotova *et al.*[70] is arguably the most relevant for understanding the loss properties of the amorphous $AlO_x$ most often employed in contemporary JJs. While some reports of amorphous $AlO_x$ have comparable or even lower $\delta_{LP}$ values than the epitaxial $Al_2O_3$ reported in this work, direct comparison of these results to one another and to the result reported here is confounded by several factors. Readers are directed to the Supporting Information for discussion of the nuances of these comparisons.

Of the ten loss measurements collected by Zotova *et al*. from LEPPCs with Al/$AlO_x$/Al capacitors, the average $\delta_{LP}$ is 1.44 × $10^{-3}$, or an approximate $F\delta^0_{\mathrm{TLS}}$ of 1.42 × $10^{-3}$. The epitaxial PLD γ-$Al_2O_3$ presented here, therefore, represents an improvement of nearly two orders of magnitude over the incumbent JJ material in terms of low-power intrinsic material loss, consistent with other reports of epitaxial $Al_2O_3$ listed in Table 1.

**Table 1.** Summary of Microwave Loss Measurements of Relevant Dielectrics from Literature, Measured in Device Geometries Involving an Integrated PPC

| Material | Reference | Deposition | Crystallinity [a] | Geometry | $\delta_{LP}$ (x $10^{-5}$) [b] | $F\delta^0_{\mathrm{TLS}}$ (x $10^{-5}$) [c] | $Q_{max}$ (x $10^5$) [d] | Area (x $10^5$ μm²) [e] |
|---|---|---|---|---|---|---|---|---|
| thick γ-$Al_2O_3$ | This work | PLD | epitaxial | LEPPC | 3.6 ± 0.3 | 3.5 ± 0.2 | 6.4 ± 0.5 | 0.244 |
| thin γ- $Al_2O_3$ | This work | PLD | epitaxial | LEPPC | 3.2 ± 0.2 | 2.8 ± 0.1 | 2.7 ± 0.1 | 0.244 |
| $Al_2O_3$ | [19] | MBE | epitaxial | LEPPC | 6 | † | † | 0.055 |
| $Al_2O_3$ | [18] | PLD | epitaxial | LEPPC | 3–5 | 2.3–4.3* | 1.4 | † |
| $Al_2O_3$ | [47] | sputter | amorphous | LEPPC | 93.7* | 92 | 0.59 | 1.225 |
| $Al_2O_3$ | [71] | ALD | amorphous | LEKID | † | † | 1.5 | 0.0952 |
| $AlO_x$ | [72] | plasma ox. | amorphous | LE overlap | 140–175 | 139–174* | 1 | 0.075–0.1155 |
| $AlO_x$ | [73] | anodic ox. | amorphous | CPWPPC | 4–22* | 4–22 | 8 | 20 |
| $AlO_x$ | [42] | e-beam evap. | amorphous | Mergemon | ≲ 0.05 | ≲ 0.047 | † | † |
| $AlO_x$ | [70] | e-beam evap. | amorphous | LEPPC | 144 | 142* | 0.45 | 0.4 |
| GaAs | [12] | MBE | epitaxial | LEPPC | 11.2* | 6.4 | 0.21 | 1.404 |
| hBN | [44] | flake transfer | single crystal | LEPPC | 0.29–0.53 | 0.14–0.38* | 2.7–6.8 | 6.25 |
| hBN | [43] | flake transfer | single crystal | vdW transmon | 2.8 | † | † | 45 |
| $Nb_2O_5$ | [74] | anodic ox. | amorphous | LEPPC | † | † | ≤ 0.01 | 7 |
| Si | [75] | sputter | amorphous | LEPPC | 150–200 | 140–190* | 0.1 | 1–3 |
| Si | [76] | PECVD | amorphous | LEPPC | 15–50 | † | † | 1.8 |
| Si | [77] | ICPCVD | amorphous | LEPPC | 5.3–9.3* | 4–8 | 0.8 | 3.9 |
| Si | [78] | Si wafer | epitaxial | LEIDC w/ fins | 0.56–1.2 | † | † | 1.75 |
| Si | [79] | SOI wafer | single crystal | LEPPC | 0.54 | 0.43* | 9 | 3 |
| Si:H | [80] | † | amorphous | LEPPC | † | F × (2.2–2.5) | 9.1 | † |
| SiC:H | [81] | PECVD | amorphous | LEPPC | † | F × (2–6) | 5 | 0.3 |
| $Si_3N_4$ | [77] | ICPCVD | amorphous | LEPPC | 98–111* | 97–110 | 1.3 | 3.9 |
| $Si_3N_4$ | [82] | LPCVD | amorphous | MKID | † | 200 | † | 62.5 |
| $SiN_x$:H | [83] | ICPCVD | amorphous | LEPPC | † | F × (2.5–120) | 2 | 0.252 |
| $SiN_x$ | [6] | CVD | amorphous | LEPPC | 20 | 19* | 1.25 | † |
| $SiN_x$ | [80] | † | amorphous | LEPPC / CPW | 10–20 | 10–20* | 5.3 | † |
| $SiN_x$ | [75] | ECR-PECVD | amorphous | LEPPC | 40–50 | 40–50* | 5.9 | 1–3 |
| $SiN_x$ | [74] | PECVD | amorphous | LEPPC | † | † | 0.033–0.1 | 7 |
| $SiN_x$ | [84] | PECVD | amorphous | LEPPC | † | F × 78 | † | 0.7 |
| $SiN_x$ | [77] | ICPCVD | amorphous | LEPPC | 4.5–16* | 4–15 | 2.0 | 3.9 |
| $SiN_x$ | [85] | LPCVD | amorphous | MKID | 183–218* | 180–215 | 0.3 | † |
| SiO | [74] | thermal evap. | amorphous | LEPPC | † | † | 0.02–0.05 | 7 |
| SiO | [86] | thermal evap. | amorphous | LEPPC | 9.5–37 | 3.6–31* | 0.17 | † |
| $SiO_2$ | [6] | CVD | amorphous | LEPPC | 500 | 494* | 0.17 | † |
| $SiO_2$ | [75] | ECR-PECVD | amorphous | LEPPC | 600 | 599* | 1 | 1–3 |

[a] Assumed to be amorphous, unless explicitly stated and shown otherwise.
[b] Low-power intrinsic loss.
[c] Intrinsic TLS loss.
[d] Maximum intrinsic quality factor.
[e] *Approximate* footprint of a single device in its totality, including both the capacitive and inductive elements.
[*] Estimated based on the data reported by the reference, such as extracting $F\delta^0_{\mathrm{TLS}}$ from $\delta_{\mathrm{LP}}$ measurements and provided S-curves, or vice versa.
[†] Items for which no comparable data were reported (*e.g.*, having large deviations of measured power or temperature from our work).

## CONCLUSIONS

In this work, we have presented an epitaxial TiN/γ-$Al_2O_3$/TiN trilayer materials stack. The materials and device preparation process starts with the PLD heteroepitaxy of TiN on sapphire. By baking and titanium-gettering the PLD chamber to reduce the water and oxygen content prior to deposition, we deposit near-stoichiometric TiN. XPS depth profiling, EDS mapping, and ToF-SIMS depth profiling reveal a low concentration of oxygen and $Ti_xO_yN_z$ species in the bulk of the TiN. Electron and X-ray diffraction of the TiN confirm excellent crystallinity and epitaxy with the sapphire substrate. Furthermore, the measured $T_c$ of our TiN at 4.8 K (Figure S2) is consistent with the highest reported $T_c$ of PLD-grown TiN on sapphire.[45] Since oxygen content in the superconductor is not a limiting factor to coherence so long as its transport properties are favorable,[87] despite residual oxygen present in the bulk, these materials properties make the initial TiN film an ideal platform for the further deposition of the trilayer by PLD.

Multimodal materials characterization confirms the growth of a cubic $Fd\bar{3}m$ γ-phase of $Al_2O_3$ and heteroepitaxy throughout the stack. Each layer in the trilayer exhibits high crystallinity, with a consistent epitaxial relationship and sharp interfaces throughout; these material qualities have proven to be critical for suppressing decoherence in heterostructures used in quantum device applications.[88] In particular, minimal structural and chemical disorder is observed in the crystalline, stoichiometric dielectric. While some oxidation of the TiN is present at the superconductor-dielectric interface, the $Ti_xO_yN_z$ interlayer thicknesses are limited to 1 – 2 nm, and there is negligible diffusion of oxygen from the γ-$Al_2O_3$ layer into the TiN bulk, demonstrating the diffusion-barrier properties of TiN.

By integrating this trilayer into LEPPC devices, we enable measurements of the microwave loss properties of γ-$Al_2O_3$. The intrinsic material TLS loss of γ-$Al_2O_3$ is found to be as low as $\delta^0_{\mathrm{TLS}} = (2.8 \pm 0.1) \times 10^{-5}$, representing an improvement of two orders of magnitude compared to the amorphous $AlO_x$ used in state-of-the-art JJs. It is possible that further improvements in microwave properties can be achieved through a better understanding of the complex structure of γ-$Al_2O_3$, through theoretical studies of intrinsic TLSs present in its spinel-like cubic crystal structure. While a quantitative estimate of qubit $T_1$ cannot be made without specifying the qubit geometry and dielectric participation, the measured resonator quality factors should not be interpreted as the intrinsic coherence limit of qubits fabricated from this materials platform. The large dielectric volume of the present LEPPC resonators places them in a TLS bath-limited regime, whereas a qubit would employ a much thinner tunnel barrier and substantially smaller capacitor area, greatly reducing both dielectric participation and the number of active TLSs[42]. Consequently, although direct qubit measurements will ultimately be required to quantify the achievable coherence, this materials system is expected to provide a favorable materials platform for compact superconducting qubit architectures, such as merged-element transmon devices.[41, 42] In addition, the low loss at high microwave powers further suggests that this materials system can benefit additional SC device applications such as microwave kinetic inductance (MKID) devices,[71, 89] hybrid superconductor-semiconductor light sources,[90] and other compact quantum circuits.[91]

## METHODS

**Pulsed Laser Deposition (PLD), Reflection High-Energy Electron Diffraction (RHEED)**

PLD of the TiN and $Al_2O_3$ films was performed using a 248 nm, 20 Hz pulsed laser with a pulse energy of 340 mJ, and a spot size of $1 \times 2$ mm$^2$. The 2-inch α-$Al_2O_3$ (001) substrates were first annealed in air at 1100°C for 2 hours. The deposition of each layer was performed at 750°C, at a background pressure for the TiN of $3 \times 10^{-4}$ Torr $N_2$ and at the chamber base pressure for the $Al_2O_3$. Ablation of a TiN target with an ultraviolet laser can cause incomplete ablation and the ejection of particles onto the wafer. It was found that this process is mitigated in the first 40,000 pulses of a freshly polished target, so multiple freshly polished TiN targets on a carousel were switched to achieve the desired thickness, free of particles. These films were grown in high-vacuum chambers typically used for PLD of oxides, which poses a challenge for oxygen-free TiN growth. The chamber was baked by operating the substrate heater at 900°C for 20 hours while ablating a titanium target at 20 Hz and 200 mJ to getter $O_2$ and $H_2O$. The residual gas during this process was tracked with a residual gas analyzer. After baking, the substrate was introduced into the deposition chamber *via* the load-lock. This process is based on the work of Torgovkin *et al*.[45] Without this baking process, the O content of TiN was 25 at%, and we could not measure a SC transition.

**X-Ray Reflectivity (XRR), High-Resolution X-Ray Diffraction (HRXRD)**

XRR and HRXRD were performed on a Rigaku SmartLab diffractometer using monochromated (double-bounce Ge (220)) Cu $K\alpha_1$ radiation. XRR was fit using the Python-based refnx analysis package.[92]

**Time-of-Flight Secondary Ion Mass Spectrometry (ToF-SIMS)**

ToF-SIMS data were collected on a blanket trilayer sample with thick $\gamma$-$Al_2O_3$ using a dual beam IONTOF M6. Negatively charged secondary ions were collected with a $Bi^+$ ion beam produced by a liquid metal ion gun (LMIG) operated at 30 keV, with an analysis area of 49 × 49 μm$^2$. Concurrently with the LMIG, a $Cs^+$ ion gun was directed at the sample, operated at 0.5 keV, with a sputtering area of 200 × 200 μm$^2$, to obtain a depth profile. SurfaceLab7 software was used to plot and analyze the SIMS data.

**Scanning Transmission Electron Microscopy (STEM), Energy-Dispersive X-Ray Spectroscopy (EDS), Electron Energy-Loss Spectroscopy (EELS), Four-Dimensional STEM (4D-STEM)**

Cross-sectional lamella samples were prepared by $Xe^+$ plasma focused ion beam (FIB) using the Thermo Scientific Helios 5 Hydra CX DualBeam Plasma FIB/SEM. Bulk-out, lift-out, and thinning processes were performed using 30 kV $Xe^+$ ion beam. Final surface cleaning steps were performed at 8 kV and 5 kV. Samples were then polished using Fischione Model 1040 NanoMill at 1 – 0.5 kV and plasma-cleaned using South Bay Technology PC-2000 Plasma Cleaner with $Ar^+$ plasma at 20 – 40 W RF power for 15 s at 150 mTorr to remove residual contamination.

STEM data were collected on an aberration-corrected JEOL JEM-ARM200CF S/TEM operating at 200 kV. The STEM convergence angle, high-angle annular dark field (HAADF) collection angle, and ABF collection angle were 27 mrad, 90 – 370 mrad, and 23 mrad, respectively. EDS data were collected in the same microscope at 200 kV with Thermo Fisher dual silicon-drift EDS detectors (1.7 steradians). EELS and 4D-STEM data were collected on a JEOL JEM-ARM300F GrandARM S/TEM operating at 300 kV with the Gatan GIF Continuum filter with K3 IS direct electron detector. The EELS collection camera length was 2 cm, with a high-resolution aperture of 2.5 mm and an acquisition step size of 3 Å per pixel. The energy dispersion was set to 90 meV/Ch to ensure the high energy resolution necessary for probing small peak shifts and splitting. A wide energy slit of 500 eV width was inserted to record the inner-shell edges in the core-loss spectrum while limiting spectrometer aberrations. 4D-STEM convergence angle was set to be 0.5 – 1.5 mrad, with a collection camera length of 4 cm. A narrow energy slit of 20 eV width was inserted to minimize diffuse inelastic scattering to maximize contrast in the collected nanobeam diffraction. A step size of 5 Å was selected to ensure high spatial resolution while avoiding oversampling. Data processing (background subtraction, denoising, signal mapping and analysis, plural scattering removal, virtual imaging, Bragg disk detection, probe calibration, geometric phase analysis for strain mapping) was performed using the Gatan Microscopy Suite software.

**X-Ray Photoelectron Spectroscopy (XPS)**

XPS data were collected with a Thermo Scientific ESCALAB 250Xi system equipped with a monochromated Al Kα source at 1486.6 eV. The XPS measurement area was ~500 × 500 $\mu m^2$. During the measurement, a flood gun was operated for charge compensation. Additionally, an $Ar^+$ ion beam was used for depth profiling, operated at 3 kV, with 180 s etch time per etch level, and 2.5 × 2.5 $mm^2$ etch area. The data analysis was performed with CasaXPS, and the data were charge-corrected to the adventitious C 1s peak at 284.8 eV.

**Device Design, Finite Element Modeling (FEM)**

Each device is patterned into a TiN/γ-$Al_2O_3$/TiN PPC, shunted by an inductor made from the bottom TiN layer, forming a quasi-lumped-element resonator. From prior measurements of the oxide thickness, capacitor and inductor dimensions are carefully designed and simulated with Ansys HFSS to obtain resonance frequencies in the appropriate measurement range (4 – 8 GHz). Following the procedure described in McRae *et al.*,[47] we perform simulations to determine the shunting capacitance $C_L$ and inductance $L$ arising from the shunting inductor. The PPC area is then selected to achieve a capacitance $C_C$ and resulting angular frequency $\omega = 1/\sqrt{L(C_C + C_L)}$. To directly measure the dielectric loss coming from the γ-$Al_2O_3$ layer, we minimize the shunting capacitance such that the electric participation ratio in the capacitor, $C_C/(C_C + C_L)$, is over 0.99 in all devices (Figure 5d). Assuming the inductor effective TLS loss tangent $F_L \tan \delta_L$ is upper bounded by $10^{-4}$, we estimate a systematic error in the extracted γ-$Al_2O_3$ loss of under 2% from misattributing shunting capacitance loss to γ-$Al_2O_3$. This bound is motivated by the measured average TLS loss tangent of $(5.8 \pm 0.4) \times 10^{-6}$ for CPW resonators patterned on the same chip as the aforementioned LEPPC devices (Figure 6b). Another potential source of systematic error in extracting the TLS loss tangent of bulk γ-$Al_2O_3$ is the presence of thin $Ti_xO_yN_z$ interlayers at the TiN/γ-$Al_2O_3$ interfaces. To estimate their potential impact, we performed two-dimensional dielectric participation simulations of the capacitor region using Ansys Maxwell 2D. For both the thin and thick γ-$Al_2O_3$ devices, we modeled 2 nm of $Ti_xO_yN_z$ at each interface, assuming a dielectric constant of 10, equal to that of γ-$Al_2O_3$. The simulated bulk γ-$Al_2O_3$ participation ratios were in excellent agreement with $p_{\text{bulk}} = 1 - \frac{t_{\text{Ti}_\text{x}\text{O}_\text{y}\text{N}_\text{z}}}{t_{\text{total}}}$, yielding 70.3% and 93.1% for the thin and thick devices, respectively. Here, $t_{\text{Ti}_\text{x}\text{O}_\text{y}\text{N}_\text{z}} = 4$ nm is the total thickness of the $Ti_xO_yN_z$ interlayers, and $t_{\text{total}}$ is the total dielectric thickness of the capacitor, equal to 13.5 nm and 58.3 nm, respectively. If the $Ti_xO_yN_z$ interlayers are assumed to be dielectric and contribute independently to TLS loss, the measured losses yield $\delta_{Al_2O_3} = 3.7 \times 10^{-5}$ and $\delta_{Ti_xO_yN_z} = 6.4 \times 10^{-6}$, corresponding to an approximately 30% higher estimate of the bulk γ-$Al_2O_3$ loss tangent. However, given the uncertainty in the electrical properties of the $Ti_xO_yN_z$ interlayers and the limited number of samples, we interpret this result only as a sensitivity analysis rather than a direct determination of their loss tangent.

**Lumped-Element Resonator with Parallel-Plate Capacitor (LEPPC) and Airbridge Fabrication**

The LEPPC and airbridge fabrication procedure involves four rounds of photolithography using the optical mask shown in Figure S3. For each round, the sample is spun coated with positive-tone photoresist (MICROPOSIT™ S1800™ series) and patterned by exposing with the Heidelberg µPG 501 maskless aligner and developing in AZ917 MIF. TiN layers are dry etched with fluorine chemistry using a SAMCO 10NR reactive ion etcher (RIE), while the γ-$Al_2O_3$ layer is physically removed using an $Ar^+$ ion mill in the Plassys BESTEK MEB550SL4-UHV double-angle electron-beam evaporator, which is also used to deposit Al for the airbridges. Photoresists are stripped using N-methyl-2-pyrrolidone (NMP) at 70°C followed by rinses in acetone, isopropyl alcohol, and deionized water. Lastly, the samples are diced with Advanced Dicing Technologies 7122 wafer dicer and packaged using the Questar Q7800 automatic wedge bonder equipped with Al wire. Detailed device fabrication procedure can be found in the Supporting Information.

**Dilution Refrigerator (DR) Measurement**

Devices were measured in a cryogen-free dilution refrigerator with a base temperature of approximately 10 mK. Samples were mounted in gold-plated copper enclosures, wire-bonded, and thermally anchored to the mixing chamber plate. The input microwave line included multiple stages of cryogenic attenuation and low-pass filtering to suppress thermal radiation and spurious high-frequency noise. The output line incorporated cryogenic isolators and a low-noise high-electron-mobility transistor amplifier at the 4 K stage, followed by additional amplification at room temperature. Scattering parameters were acquired using a vector network analyzer. This configuration enables quantitative measurements of SC resonators in the 4 – 8 GHz range down to excitation powers corresponding to sub-single-photon intracavity occupation. More details on the wiring configuration are given in Vallières *et al*.[93]

For each device, the complex resonator response was recorded as a function of input power at a fixed temperature. Individual resonance traces were analyzed using the diameter correction method, which accounts for systematic effects such as impedance mismatches, cable delays, and asymmetries in the microwave environment.[94, 95] This analysis yields the resonance frequency, internal quality factor $Q_i$, coupling quality factor $Q_c$, and an asymmetry parameter characterizing the impedance mismatch.

To quantify dielectric loss, the extracted internal quality factor was analyzed as a function of microwave drive power (Figure 6) using a TLS loss model.[62] This approach separates the saturable TLS contribution from residual power-independent losses, providing an effective TLS loss tangent for each resonator. From the fitted TLS parameters, we place an upper bound on the intrinsic loss tangent of the γ-$Al_2O_3$ dielectric.

**Scanning Electron Microscopy (SEM)**

Top-down (0º) and tilted (45º) SEM imaging was performed at 3 kV using the Hitachi SU8700 UHR FEG-SEM. The images were collected using a combination of the in-chamber Everhart–Thornley detector collecting secondary electrons and an in-column detector collecting backscattered electrons.

**Superconducting Transition Temperature ($T_c$) Characterization**

Electrical transport measurements on the PLD TiN were performed using a Quantum Design Dynacool physical properties measurement system (PPMS) in a four-probe geometry. The sample was wire-bonded to the sample puck using a home-built In-Au wire bonder.

## AUTHOR INFORMATION

**Corresponding Authors**

[*]Vinayak P. Dravid – Email: v-dravid@northwestern.edu

[*]Mark C. Hersam – Email: m-hersam@northwestern.edu

[*]Michael J. Bedzyk – Email: bedzyk@northwestern.edu

**Present Addresses**

[†]David A. Garcia-Wetten – Department of Physics and Applied Physics, Stanford University, Stanford, CA 94305, USA; SLAC National Accelerator Laboratory, Menlo Park, CA 94025, USA

[†]Mitchell J. Walker – HZO Inc., Morrisville, NC 27560, USA

**Author Contributions**

[‡]D.A.G.-W., M.J.W., P.G.L., and A.V. contributed equally to this work; D.A.G.-W. and M.J.W. conceived the project with contributions from P.G.L. and A.V.; D.A.G.-W., M.J.W. and P.G.L. designed the experiments for materials analysis; D.A.G.-W. deposited the materials stack using PLD with contributions from M.G.J.-G; D.A.G.-W. performed X-ray characterization; P.G.L. performed electron microscopy characterization with contributions from M.A.A.; M.J.W. performed XPS, ToF-SIMS, and PPMS characterization with contributions from D.A.G.-W. and

D.P.G.; A.V. designed the devices and performed microwave measurements in the DR.; M.J.W. fabricated the devices with contributions from D.A.G.-W. and P.G.L.; D.A.G.-W., M.J.W. and P.G.L. wrote the manuscript with input from all other authors.; D.P.G., A.G., J.K., V.P.D., M.C.H. and M.J.B. supervised the project; All authors have given approval to the final version of the manuscript.

**Notes**

The authors declare no competing interests.

**ACKNOWLEDGMENTS**

The authors thank G. dos Santos and R. dos Reis for discussions on 4D-STEM data analysis, D. B. Buchholz for assistance with PLD process design, and Y. Lu and C. R. McRae for discussions on device design and microwave measurements.

This work was primarily supported by the U.S. Department of Energy, Office of Science, National Quantum Information Science Research Centers, Superconducting Quantum Materials and Systems Center (SQMS), under Contract No. 89243024CSC000002. Fermilab is operated by Fermi Forward Discovery Group, LLC under Contract No. 89243024CSC000002 with the U.S. Department of Energy, Office of Science, Office of High Energy Physics. This work made use of the EPIC (RRID: SCR_026361), Keck-II (RRID: SCR_026360), and NUFAB (RRID: SCR_017779) facilities of Northwestern University's NU*ANCE* Center, which has received support from the IIN and Northwestern's MRSEC program (NSF DMR-2308691). Research reported in this publication was supported in part by instrumentation provided by the Office of The Director, National Institutes of Health of the National Institutes of Health under Award Number S10OD026871. The content is solely the responsibility of the authors and does not necessarily represent the official views of the National Institutes of Health. This work made use of the NU XRD Facility (RRID: SCR_017866) and the Pulsed Laser Deposition Facility (RRID: SCR_017889), supported by the MRSEC program (NSF-DMR-2308691) and SHyNE (NSF ECCS-2025633). This work made use of the MLTOF Facility which has received support from the MRSEC Program (NSF DMR-2308691), the Center for Hierarchical Materials Design, and Northwestern University. This work made use of the Pritzker Nanofabrication Facility, part of the Pritzker School of Molecular Engineering at the University of Chicago, which receives support from SHyNE (NSF ECCS-2025633), a node of the NSF National Nanotechnology Coordinated Infrastructure (RRID: SCR_022955).

## ASSOCIATED CONTENT

An earlier version of this article have been previously submitted to the arXiv pre-print server: Garcia-Wetten, D. A.; Walker, M. J.; Lim, P. G.; Vallières, A.; Jimenez-Guillermo, M. G.; Alvarado, M. A.; Goronzy, D. P.; Grassellino, A.; Koch, J.; Dravid, V. P.; Hersam, M. C.; Bedzyk, M. J. Oxide-nitride heteroepitaxy for low-loss dielectrics in superconducting quantum circuits. 2026, 2603.29065. arXiv.org e-Print archive. https://doi.org/10.48550/arXiv.2603.29065 (accessed September 15, 2026).

The data underlying this study are available in the published article and its Supporting Information.

**Supporting Information**

"Additional experimental details and methods, including 4D-STEM, $T_c$ measurements, detailed fabrication procedure, photomask pattern, LEPPC-airbridge interface characterization, temperature sweeps of microwave measurements, and an extended discussion on Table 1 (PDF)".

## ABBREVIATIONS

SC, superconducting; QIS, quantum information science; JJ, Josephson junction; TLS, two-level system; CPW, coplanar waveguide; PLD, pulsed laser deposition; XRR, X-ray reflectivity; XRD, X-ray diffraction; ToF-SIMS, time-of-flight secondary ion mass spectrometry; STEM, scanning transmission electron microscopy; XPS, X-ray photoelectron spectroscopy; EELS, electron energy-loss spectroscopy; PPC, parallel-plate capacitor; EDS, energy-dispersive X-ray spectroscopy; RHEED, reflectance high-energy electron diffraction; HAADF, high-angle annular dark field; ABF, annular bright field; FFT, fast Fourier transform; LEPPC, lumped-element resonator with parallel-plate capacitor; SEM, scanning electron microscopy; FEM, finite element modeling; MBE, molecular-beam epitaxy; ALD, atomic layer deposition; PECVD, plasma-enhanced chemical vapor deposition; ICPCVD, inductively coupled plasma chemical vapor deposition; SOI, silicon on insulator; LPCVD, low pressure chemical vapor deposition; ECR, electron cyclotron resonance; LEKID, lumped-element kinetic inductance detector; MKID, microwave kinetic inductance detector; LEIDC, lumped-element interdigitated capacitor; LMIG, liquid metal ion gun; FIB, focused ion beam; RIE, reactive ion etcher; RF, radio frequency; NMP, n-methyl-2-pyrrolidone; DR, dilution refrigerator.

## TABLE OF CONTENTS GRAPHIC

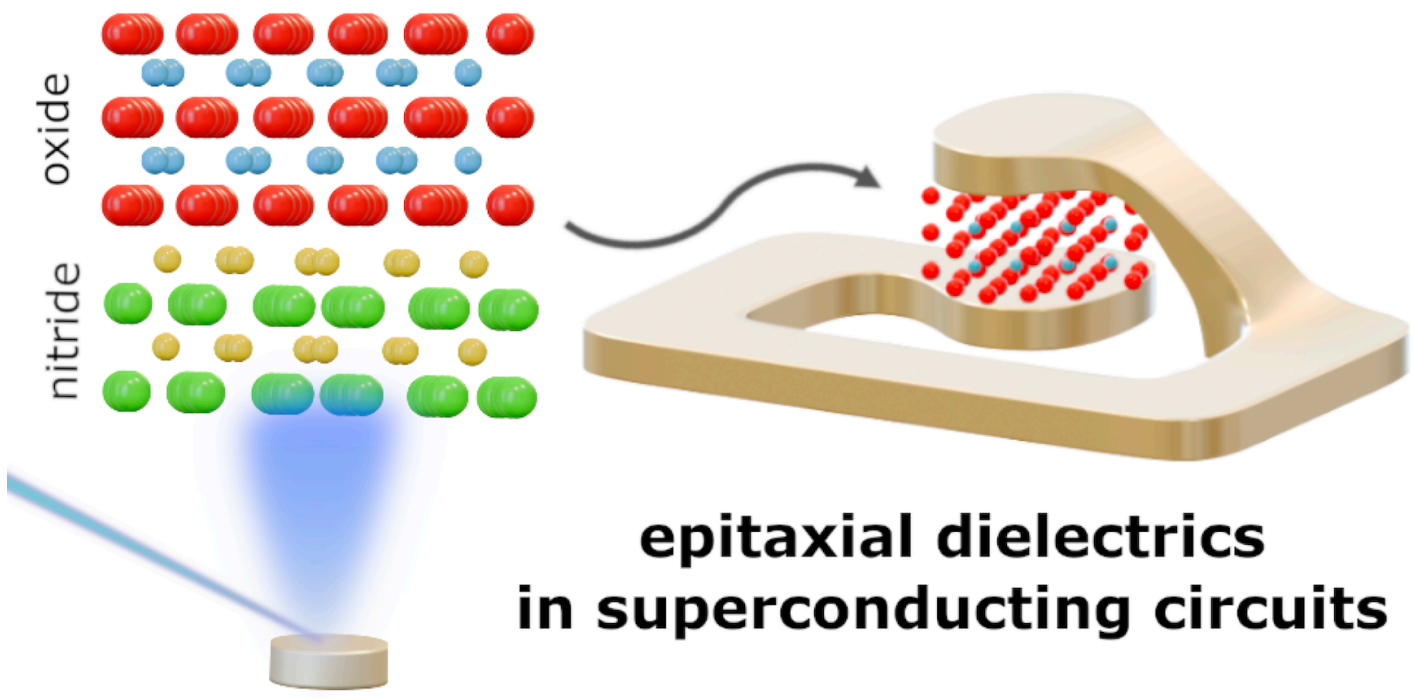

## REFERENCES

(1) Shnirman, A.; Schön, G.; Hermon, Z. Quantum Manipulations of Small Josephson Junctions. *Phys. Rev. Lett.* **1997**, *79*, 2371-2374.
(2) Bouchiat, V.; Vion, D.; Joyez, P.; Esteve, D.; Devoret, M. H. Quantum Coherence with a Single Cooper Pair. *Phys. Scr.* **1998**, *1998*, 165.
(3) Nakamura, Y.; Pashkin, Y. A.; Tsai, J. S. Coherent Control of Macroscopic Quantum States in a Single-Cooper-Pair Box. *Nature* **1999**, *398*, 786-788.
(4) Lisenfeld, J.; Grabovskij, G. J.; Müller, C.; Cole, J. H.; Weiss, G.; Ustinov, A. V. Observation of Directly Interacting Coherent Two-Level Systems in an Amorphous Material. *Nat. Commun.* **2015**, *6*, 6182.
(5) Müller, C.; Cole, J. H.; Lisenfeld, J. Towards Understanding Two-Level-Systems in Amorphous Solids: Insights from Quantum Circuits. *Rep. Prog. Phys.* **2019**, *82*, 124501.
(6) Martinis, J. M.; Cooper, K. B.; McDermott, R.; Steffen, M.; Ansmann, M.; Osborn, K. D.; Cicak, K.; Oh, S.; Pappas, D. P.; Simmonds, R. W.; Yu, C. C. Decoherence in Josephson Qubits from Dielectric Loss. *Phys. Rev. Lett.* **2005**, *95*, 210503.
(7) Lisenfeld, J.; Bilmes, A.; Matityahu, S.; Zanker, S.; Marthaler, M.; Schechter, M.; Schon, G.; Shnirman, A.; Weiss, G.; Ustinov, A. V. Decoherence Spectroscopy with Individual Two-Level Tunneling Defects. *Sci. Rep.* **2016**, *6*, 23786.
(8) Burnett, J. J.; Bengtsson, A.; Scigliuzzo, M.; Niepce, D.; Kudra, M.; Delsing, P.; Bylander, J. Decoherence Benchmarking of Superconducting Qubits. *npj Quantum Inf.* **2019**, *5*, 54.
(9) Zeng, L.; Tran, D. T.; Tai, C. W.; Svensson, G.; Olsson, E. Atomic Structure and Oxygen Deficiency of the Ultrathin Aluminium Oxide Barrier in Al/Alox/Al Josephson Junctions. *Sci. Rep.* **2016**, *6*, 29679.
(10) Oliver, W. D.; Welander, P. B. Materials in Superconducting Quantum Bits. *MRS Bull.* **2013**, *38*, 816-825.
(11) McFadden, A. P.; Goswami, A.; Seas, M.; McRae, C. R. H.; Zhao, R.; Pappas, D. P.; Palmstrøm, C. J. Epitaxial Al/Gaas/Al Tri-Layers Fabricated Using a Novel Wafer-Bonding Technique. *J. Appl. Phys.* **2020**, *128*, 115301.
(12) McRae, C. R. H.; McFadden, A.; Zhao, R.; Wang, H.; Long, J. L.; Zhao, T.; Park, S.; Bal, M.; Palmstrøm, C. J.; Pappas, D. P. Cryogenic Microwave Loss in Epitaxial Al/Gaas/Al Trilayers for Superconducting Circuits. *J. Appl. Phys.* **2021**, *129*, 025109.
(13) Nakamura, Y.; Terai, H.; Inomata, K.; Yamamoto, T.; Qiu, W.; Wang, Z. Superconducting Qubits Consisting of Epitaxially Grown Nbn/Aln/Nbn Josephson Junctions. *Appl. Phys. Lett.* **2011**, *99*, 212502.
(14) Wang, D.; Wu, Y.; Pieczulewski, N.; Garg, P.; Pace, M. C. C.; Bøttcher, C. G. L.; Mazumder, B.; Muller, D. A.; Tang, H. X. All-Nitride Superconducting Qubits Based on Atomic Layer Deposition. *Nat. Mater.* **2026**.
(15) Fricke, K. Piezoelectric Properties of Gaas for Application in Stress Transducers. *J. Appl. Phys.* **1991**, *70*, 914-918.
(16) Dubois, M.-A.; Muralt, P. Properties of Aluminum Nitride Thin Films for Piezoelectric Transducers and Microwave Filter Applications. *Appl. Phys. Lett.* **1999**, *74*, 3032-3034.
(17) Welander, P. B. Epitaxial Aluminum Oxide Thin Films on Niobium (110): A Study of Their Growth and Their Use in Superconducting Tunnel-Junctions. PhD Thesis, University of Illinois at Urbana-Champaign, 2007.

(18) Cho, K. H.; Patel, U.; Podkaminer, J.; Gao, Y.; Folkman, C. M.; Bark, C. W.; Lee, S.; Zhang, Y.; Pan, X. Q.; McDermott, R.; Eom, C. B. Epitaxial Al2o3 Capacitors for Low Microwave Loss Superconducting Quantum Circuits. *APL Mater.* **2013**, *1*, 042115.
(19) Weides, M. P.; Kline, J. S.; Vissers, M. R.; Sandberg, M. O.; Wisbey, D. S.; Johnson, B. R.; Ohki, T. A.; Pappas, D. P. Coherence in a Transmon Qubit with Epitaxial Tunnel Junctions. *Appl. Phys. Lett.* **2011**, *99*, 262502.
(20) Kline, J. S.; Vissers, M. R.; Da Silva, F. C. S.; Wisbey, D. S.; Weides, M.; Weir, T. J.; Turek, B.; Braje, D. A.; Oliver, W. D.; Shalibo, Y.; Katz, N.; Johnson, B. R.; Ohki, T. A.; Pappas, D. P. Sub-Micrometer Epitaxial Josephson Junctions for Quantum Circuits. *Supercon. Sci. Technol.* **2012**, *25*, 025005.
(21) Hulm, J. K.; Goodman, B. B. Superconducting Properties of Rhenium, Ruthenium, and Osmium. *Phys. Rev.* **1957**, *106*, 659-671.
(22) Jena, D.; Page, R.; Casamento, J.; Dang, P.; Singhal, J.; Zhang, Z.; Wright, J.; Khalsa, G.; Cho, Y.; Xing, H. G. The New Nitrides: Layered, Ferroelectric, Magnetic, Metallic and Superconducting Nitrides to Boost the Gan Photonics and Electronics Eco-System. *Jpn. J. Appl. Phys.* **2019**, *58*, SC0801.
(23) Dang, P.; Khalsa, G.; Chang, C. S.; Katzer, D. S.; Nepal, N.; Downey, B. P.; Wheeler, V. D.; Suslov, A.; Xie, A.; Beam, E.; Cao, Y.; Lee, C.; Muller, D. A.; Xing, H. G.; Meyer, D. J.; Jena, D. An All-Epitaxial Nitride Heterostructure with Concurrent Quantum Hall Effect and Superconductivity. *Sci. Adv.* **2021**, *7*, eabf1388.
(24) Ithepalli, A.; Rajapurohita, A. R.; Singh, A.; Singh, R.; Wright, J.; Rana, F.; Fatemi, V.; Xing, H.; Jena, D. Structural, Electronic, and Superconducting Properties of Mbe-Grown Tantalum Nitride Films on C-Plane Sapphire. *Appl. Phys. Lett.* **2025**, *126*, 222601.
(25) Foshat, P.; Poorgholam-Khanjari, S.; Seferai, V.; Feng, H.; Johny, S.; Mukhanov, O. A.; Hutchings, M.; Hadfield, R. H.; Weides, M.; Delfanazari, K. Quasiparticle Dynamics in Niobium Nitride Superconducting Microwave Resonators at Single-Photon Regime. *IEEE Transactions on Quantum Engineering* **2025**, *6*, 3102608-3102608.
(26) Foshat, P.; Baity, P.; Danilin, S.; Seferai, V.; Poorgholam-Khanjari, S.; Feng, H.; Mukhanov, O. A.; Hutchings, M.; Hadfield, R. H.; Weides, M.; Delfanazari, K. Characterizing Niobium Nitride-Based Superconducting Coplanar Waveguide Resonators for Microwave Hybrid Circuit Quantum Electrodynamics. *IEEE Trans. Appl. Supercond.* **2025**, *35*, 0601116-0601116.
(27) Sarker, J.; Garg, P.; Zhu, M.; Rouleau, C. M.; Hwang, J.; Osei-Agyemang, E.; Mazumder, B. Understanding the Structural–Chemical Evolution of Epitaxial Nbn/Al2o3/Nbn Trilayers with Varying Nbn Thickness. *ACS Appl. Eng. Mater. ACS Appl Eng Mater* **2023**, *1*, 3227-3236.
(28) Vissers, M. R.; Gao, J.; Wisbey, D. S.; Hite, D. A.; Tsuei, C. C.; Corcoles, A. D.; Steffen, M.; Pappas, D. P. Low Loss Superconducting Titanium Nitride Coplanar Waveguide Resonators. *Appl. Phys. Lett.* **2010**, *97*, 232509.
(29) Chang, J. B.; Vissers, M. R.; Córcoles, A. D.; Sandberg, M.; Gao, J.; Abraham, D. W.; Chow, J. M.; Gambetta, J. M.; Beth Rothwell, M.; Keefe, G. A.; Steffen, M.; Pappas, D. P. Improved Superconducting Qubit Coherence Using Titanium Nitride. *Appl. Phys. Lett.* **2013**, *103*, 012602-012602.
(30) Melville, A.; Calusine, G.; Woods, W.; Serniak, K.; Golden, E.; Niedzielski, B. M.; Kim, D. K.; Sevi, A.; Yoder, J. L.; Dauler, E. A.; Oliver, W. D. Comparison of Dielectric Loss in Titanium Nitride and Aluminum Superconducting Resonators. *Appl. Phys. Lett.* **2020**, *117*, 124004.

(31) Torgovkin, A.; Ruhtinas, A.; Maasilta, I. J. Normal Metal-Insulator-Superconductor Tunnel Junctions with Pulsed Laser Deposited Titanium Nitride as Superconductor. *IEEE Trans. Appl. Supercond.* **2021**, *31*, 1-4.
(32) Gao, R.; Yu, W.; Deng, H.; Ku, H. S.; Li, Z.; Wang, M.; Miao, X.; Lin, Y.; Deng, C. Epitaxial Titanium Nitride Microwave Resonators: Structural, Chemical, Electrical, and Microwave Properties. *Phys. Rev. Mater.* **2022**, *6*, 036202.
(33) Deng, H.; Song, Z.; Gao, R.; Xia, T.; Bao, F.; Jiang, X.; Ku, H. S.; Li, Z.; Ma, X.; Qin, J.; Sun, H.; Tang, C.; Wang, T.; Wu, F.; Yu, W.; Zhang, G.; Zhang, X.; Zhou, J.; Zhu, X.; Shi, Y.*; et al.* Titanium Nitride Film on Sapphire Substrate with Low Dielectric Loss for Superconducting Qubits. *Phys. Rev. Appl.* **2023**, *19*, 024013.
(34) Bal, M.; Murthy, A. A.; Zhu, S.; Crisa, F.; You, X.; Huang, Z.; Roy, T.; Lee, J.; Zanten, D. M. P. v.; Pilipenko, R.; Nekrashevich, I.; Bafia, D.; Krasnikova, Y.; Kopas, C. J.; Lachman, E. O.; Miller, D.; Mutus, J. Y.; Reagor, M. J.; Cansizoglu, H.; Marshall, J.*; et al.* Systematic Improvements in Transmon Qubit Coherence Enabled by Niobium Surface Encapsulation. *npj Quantum Inf.* **2024**, *10*, 1-8.
(35) Nicolet, M. A. Diffusion Barriers in Thin Films. *Thin Solid Films* **1978**, *52*, 415-443.
(36) Araujo, R. A.; Zhang, X.; Wang, H. Epitaxial Cubic Hfn Diffusion Barriers Deposited on Si (001) by Using a Tin Buffer Layer. *J. Vac. Sci. Technol., B:Microelectron. Nanometer Struct.--Process., Meas., Phenom.* **2008**, *26*, 1871-1874.
(37) Koch, J.; Yu, T. M.; Gambetta, J. M.; Houck, A. A.; Schuster, D. I.; Majer, J.; Blais, A.; Devoret, M. H.; Girvin, S. M.; Schoelkopf, R. J. Charge Insensitive Qubit Design Derived from the Cooper Pair Box. *Phys. Rev. A* **2007**, *76*, 042319.
(38) Wang, C.; Axline, C.; Gao, Y. Y.; Brecht, T.; Chu, Y.; Frunzio, L.; Devoret, M. H.; Schoelkopf, R. J. Surface Participation and Dielectric Loss in Superconducting Qubits. *Appl. Phys. Lett.* **2015**, *107*, 162601.
(39) Gambetta, J. M.; Murray, C. E.; Fung, Y. K. K.; McClure, D. T.; Dial, O.; Shanks, W.; Sleight, J. W.; Steffen, M. Investigating Surface Loss Effects in Superconducting Transmon Qubits. *IEEE Trans. Appl. Supercond.* **2017**, *27*, 1700205.
(40) Göppl, M.; Fragner, A.; Baur, M.; Bianchetti, R.; Filipp, S.; Fink, J. M.; Leek, P. J.; Puebla, G.; Steffen, L.; Wallraff, A. Coplanar Waveguide Resonators for Circuit Quantum Electrodynamics. *J. Appl. Phys.* **2008**, *104*, 113904.
(41) Zhao, R.; Park, S.; Zhao, T.; Bal, M.; McRae, C. R. H.; Long, J.; Pappas, D. P. Merged-Element Transmon. *Phys. Rev. Appl.* **2020**, *14*, 064006.
(42) Mamin, H. J.; Huang, E.; Carnevale, S.; Rettner, C. T.; Arellano, N.; Sherwood, M. H.; Kurter, C.; Trimm, B.; Sandberg, M.; Shelby, R. M.; Mueed, M. A.; Madon, B. A.; Pushp, A.; Steffen, M.; Rugar, D. Merged-Element Transmons: Design and Qubit Performance. *Phys. Rev. Appl.* **2021**, *16*, 024023.
(43) Antony, A.; Gustafsson, M. V.; Ribeill, G. J.; Ware, M.; Rajendran, A.; Govia, L. C. G.; Ohki, T. A.; Taniguchi, T.; Watanabe, K.; Hone, J.; Fong, K. C. Miniaturizing Transmon Qubits Using Van Der Waals Materials. *Nano Lett.* **2021**, *21*, 10122-10126.
(44) Wang, J. I. J.; Yamoah, M. A.; Li, Q.; Karamlou, A. H.; Dinh, T.; Kannan, B.; Braumüller, J.; Kim, D.; Melville, A. J.; Muschinske, S. E.; Niedzielski, B. M.; Serniak, K.; Sung, Y.; Winik, R.; Yoder, J. L.; Schwartz, M. E.; Watanabe, K.; Taniguchi, T.; Orlando, T. P.; Gustavsson, S.*; et al.* Hexagonal Boron Nitride as a Low-Loss Dielectric for Superconducting Quantum Circuits and Qubits. *Nat. Mater.* **2022**, *21*, 398-403.

(45) Torgovkin, A.; Chaudhuri, S.; Ruhtinas, A.; Lahtinen, M.; Sajavaara, T.; Maasilta, I. J. High Quality Superconducting Titanium Nitride Thin Film Growth Using Infrared Pulsed Laser Deposition. *Supercon. Sci. Technol.* **2018**, *31*, 055017.
(46) Balakrishnan, G.; Babu, R. V.; Shin, K. S.; Song, J. I. Growth of Highly Oriented γ- and α-Al2o3 Thin Films by Pulsed Laser Deposition. *Opt. Laser Technol.* **2014**, *56*, 317-321.
(47) McRae, C. R. H.; Lake, R. E.; Long, J. L.; Bal, M.; Wu, X.; Jugdersuren, B.; Metcalf, T. H.; Liu, X.; Pappas, D. P. Dielectric Loss Extraction for Superconducting Microwave Resonators. *Appl. Phys. Lett.* **2020**, *116*, 194003.
(48) Welander, P. B.; Bolkhovsky, V.; Weir, T. J.; Gouker, M. A.; Oliver, W. D. Shadow Evaporation of Epitaxial Al/Al2o3/Al Tunnel Junctions on Sapphire Utilizing an Inorganic Bilayer Mask. 2012, 1203.6007. arXiv.org e-Print archive. https://doi.org/10.48550/arXiv.1203.6007 (accessed February 6, 2026).
(49) Smrčok, Ľ.; Langer, V.; Křesťan, J. γ-Alumina: A Single-Crystal X-Ray Diffraction Study. *Acta Crystallogr., Sect. C:Cryst. Struct. Commun.* **2006**, *62*, i83-i84.
(50) Prins, R. On the Structure of γ-Al2o3. *Journal of Catalysis* **2020**, *392*, 336-346.
(51) Merckling, C.; El-Kazzi, M.; Saint-Girons, G.; Hollinger, G.; Largeau, L.; Patriarche, G.; Favre-Nicolin, V.; Marty, O. Growth of Crystalline γ-Al2o3 on Si by Molecular Beam Epitaxy: Influence of the Substrate Orientation. *J. Appl. Phys.* **2007**, *102*, 024101.
(52) Richardson, C. J. K.; Alexander, A.; Weddle, C. G.; Arey, B.; Olszta, M. Low-Loss Superconducting Titanium Nitride Grown Using Plasma-Assisted Molecular Beam Epitaxy. *J. Appl. Phys.* **2020**, *127*, 235302-235302.
(53) Jaeger, D.; Patscheider, J. A Complete and Self-Consistent Evaluation of Xps Spectra of Tin. *J. Electron Spectrosc. Relat. Phenom.* **2012**, *185*, 523-534.
(54) Porte, L.; Roux, L.; Hanus, J. Vacancy Effects in the X-Ray Photoelectron Spectra of Tinx. *Phys. Rev. B* **1983**, *28*, 3214-3224.
(55) Filatova, E. O.; Konashuk, A. S.; Sakhonenkov, S. S.; Sokolov, A. A.; Afanas'ev, V. V. Re-Distribution of Oxygen at the Interface between γ-Al2o3 and Tin. *Sci. Rep.* **2017**, *7*, 4541.
(56) Bertóti, I.; Mohai, M.; Sullivan, J. L.; Saied, S. O. Surface Characterisation of Plasma-Nitrided Titanium: An Xps Study. *Appl. Surf. Sci.* **1995**, *84*, 357-371.
(57) Abdallah, I.; Dupressoire, C.; Laffont, L.; Monceau, D.; Put, A. V. Stem-Eels Identification of Tioxny, Tin, Ti2n and O, N Dissolution in the Ti2642s Alloy Oxidized in Synthetic Air at 650 °C. *Corros. Sci* **2019**, *153*, 191-199.
(58) Toyoda, S.; Shinohara, T.; Kumigashira, H.; Oshima, M.; Kato, Y. Significant Increase in Conduction Band Discontinuity Due to Solid Phase Epitaxy of Al2o3 Gate Insulator Films on Gan Semiconductor. *Appl. Phys. Lett.* **2012**, *101*, 231607.
(59) Manassidis, I.; Gillan, M. J. Structure and Energetics of Alumina Surfaces Calculated from First Principles. *J. Am. Ceram. Soc.* **1994**, *77*, 335-338.
(60) Filatova, E. O.; Konashuk, A. S.; Schaefers, F.; Afanas’ev, V. V. Metallization-Induced Oxygen Deficiency of γ-Al2o3 Layers. *J. Phys. Chem. C* **2016**, *120*, 8979-8985.
(61) Chen, Z.; Megrant, A.; Kelly, J.; Barends, R.; Bochmann, J.; Chen, Y.; Chiaro, B.; Dunsworth, A.; Jeffrey, E.; Mutus, J. Y.; O'Malley, P. J. J.; Neill, C.; Roushan, P.; Sank, D.; Vainsencher, A.; Wenner, J.; White, T. C.; Cleland, A. N.; Martinis, J. M. Fabrication and Characterization of Aluminum Airbridges for Superconducting Microwave Circuits. *Appl. Phys. Lett.* **2014**, *104*, 052602.

(62) McRae, C. R. H.; Wang, H.; Gao, J.; Vissers, M. R.; Brecht, T.; Dunsworth, A.; Pappas, D. P.; Mutus, J. Materials Loss Measurements Using Superconducting Microwave Resonators. *Rev. Sci. Instrum.* **2020**, *91*, 091101.
(63) Martin, N.; Banakh, O.; Santo, A. M. E.; Springer, S.; Âs, R. S.; Takadoum, J.; Le Âvy, F. Correlation between Processing and Properties of Tioxny Thin Films Sputter Deposited by the Reactive Gas Pulsing Technique. *Appl. Surf. Sci.* **2001**, *185*, 123-123.
(64) Chappé, J. M.; Martin, N.; Lintymer, J.; Sthal, F.; Terwagne, G.; Takadoum, J. Titanium Oxynitride Thin Films Sputter Deposited by the Reactive Gas Pulsing Process. *Appl. Surf. Sci.* **2007**, *253*, 5312-5316.
(65) Akazawa, H. Sputtering Characteristics, Crystal Structures, and Transparent Conductive Properties of Tioxny Films Deposited on α-Al2o3 (0001) and Glass Substrates. *Appl. Surf. Sci.* **2012**, *263*, 307-313.
(66) Jia, L.; Lu, H.; Ran, Y.; Zhao, S.; Liu, H.; Li, Y.; Jiang, Z.; Wang, Z. Structural and Dielectric Properties of Ion Beam Deposited Titanium Oxynitride Thin Films. *Journal of Materials Science* **2019**, *54*, 1452-1461.
(67) Zhuravel, A. P.; Anlage, S. M.; Remillard, S. K.; Lukashenko, A. V.; Ustinov, A. V. Effect of Laalo3 Twin-Domain Topology on Local Dc and Microwave Properties of Cuprate Films. *J. Appl. Phys.* **2010**, *108*, 033920-033920.
(68) Arzeo, M.; Lombardi, F.; Bauch, T. Microwave Losses in Mgo, Laalo3, and (La0.3sr 0.7)(Al0.65ta0.35)O3 Dielectrics at Low Power and in the Millikelvin Temperature Range. *Appl. Phys. Lett.* **2014**, *104*, 212601-212601.
(69) Biznárová, J.; Osman, A.; Rehnman, E.; Chayanun, L.; Križan, C.; Malmberg, P.; Rommel, M.; Warren, C.; Delsing, P.; Yurgens, A.; Bylander, J.; Fadavi Roudsari, A. Mitigation of Interfacial Dielectric Loss in Aluminum-on-Silicon Superconducting Qubits. *npj Quantum Inf.* **2024**, *10*, 78-78.
(70) Zotova, J.; Wang, R.; Semenov, A.; Zhou, Y.; Khrapach, I.; Tomonaga, A.; Astafiev, O.; Tsai, J.-S. Compact Superconducting Microwave Resonators Based on Al-Alox-Al Capacitors. *Phys. Rev. Appl.* **2023**, *19*, 044067.
(71) Beldi, S.; Boussaha, F.; Hu, J.; Monfardini, A.; Traini, A.; Levy-Bertrand, F.; Chaumont, C.; Gonzales, M.; Firminy, J.; Reix, F.; Rosticher, M.; Mignot, S.; Piat, M.; Bonifacio, P. High Q-Factor near Infrared and Visible Al2o3-Based Parallel-Plate Capacitor Kinetic Inductance Detectors. *Opt. Express* **2019**, *27*, 13319.
(72) Deng, C.; Otto, M.; Lupascu, A. Characterization of Low-Temperature Microwave Loss of Thin Aluminum Oxide Formed by Plasma Oxidation. *Appl. Phys. Lett.* **2014**, *104*, 043506.
(73) Brehm, J. D.; Bilmes, A.; Weiss, G.; Ustinov, A. V.; Lisenfeld, J. Transmission-Line Resonators for the Study of Individual Two-Level Tunneling Systems. *Appl. Phys. Lett.* **2017**, *111*, 112601.
(74) Kaiser, C.; Skacel, S. T.; Wünsch, S.; Dolata, R.; Mackrodt, B.; Zorin, A.; Siegel, M. Measurement of Dielectric Losses in Amorphous Thin Films at Gigahertz Frequencies Using Superconducting Resonators. *Supercon. Sci. Technol.* **2010**, *23*, 075008.
(75) Cicak, K.; Li, D.; Strong, J. A.; Allman, M. S.; Altomare, F.; Sirois, A. J.; Whittaker, J. D.; Teufel, J. D.; Simmonds, R. W. Low-Loss Superconducting Resonant Circuits Using Vacuum-Gap-Based Microwave Components. *Appl. Phys. Lett.* **2010**, *96*, 093502.
(76) Lecocq, F.; Ranzani, L.; Peterson, G. A.; Cicak, K.; Simmonds, R. W.; Teufel, J. D.; Aumentado, J. Nonreciprocal Microwave Signal Processing with a Field-Programmable Josephson Amplifier. *Phys. Rev. Appl.* **2017**, *7*, 024028.

(77) Sun, J.; Shu, S.; Chai, Y.; Zhu, L.; Zhang, L.; Li, Y.; Liu, Z.; Li, Z.; Shi, W.; Xu, Y.; Yan, D.; Guo, W.; Wang, Y.; Liu, C. Cryogenic Microwave Performance of Silicon Nitride and Amorphous Silicon Deposited Using Low-Temperature Icpcvd. *J. Low Temp. Phys.* **2024**, *217*, 464-471.
(78) Goswami, A.; McFadden, A. P.; Zhao, T.; Inbar, H.; Dong, J. T.; Zhao, R.; McRae, C. R. H.; Simmonds, R. W.; Palmstrøm, C. J.; Pappas, D. P. Towards Merged-Element Transmons Using Silicon Fins: The Finmet. *Appl. Phys. Lett.* **2022**, *121*, 064001.
(79) Weber, S. J.; Murch, K. W.; Slichter, D. H.; Vijay, R.; Siddiqi, I. Single Crystal Silicon Capacitors with Low Microwave Loss in the Single Photon Regime. *Appl. Phys. Lett.* **2011**, *98*, 172510.
(80) O'Connell, A. D.; Ansmann, M.; Bialczak, R. C.; Hofheinz, M.; Katz, N.; Lucero, E.; McKenney, C.; Neeley, M.; Wang, H.; Weig, E. M.; Cleland, A. N.; Martinis, J. M. Microwave Dielectric Loss at Single Photon Energies and Millikelvin Temperatures. *Appl. Phys. Lett.* **2008**, *92*, 112903.
(81) Kouwenhoven, K.; Doorn, G. P. J. v.; Buijtendorp, B. T.; Rooij, S. A. H. d.; Lamers, D.; Thoen, D. J.; Murugesan, V.; Baselmans, J. J. A.; Visser, P. J. d. Geometry Dependence of Two-Level-System Noise and Loss in a-Sic:H Parallel-Plate Capacitors for Superconducting Microwave Resonators. *Phys. Rev. Appl.* **2024**, *21*, 044036.
(82) Hood II, J. C.; Barry, P. S.; Cecil, T.; Chang, C. L.; Li, J.; Meyer, S. S.; Pan, Z.; Shirokoff, E.; Tang, A. Testing Low-Loss Microstrip Materials with Mkids for Microwave Applications. *J. Low Temp. Phys.* **2022**, *209*, 1189-1195.
(83) Paik, H.; Osborn, K. D. Reducing Quantum-Regime Dielectric Loss of Silicon Nitride for Superconducting Quantum Circuits. *Appl. Phys. Lett.* **2010**, *96*, 072505.
(84) Sarabi, B.; Ramanayaka, A. N.; Burin, A. L.; Wellstood, F. C.; Osborn, K. D. Projected Dipole Moments of Individual Two-Level Defects Extracted Using Circuit Quantum Electrodynamics. *Phys. Rev. Lett.* **2016**, *116*, 167002.
(85) Pan, Z.; Barry, P. S.; Cecil, T.; Albert, C.; Bender, A. N.; Chang, C. L.; Gualtieri, R.; Hood, J.; Li, J.; Zhang, J.; Lisovenko, M.; Novosad, V.; Wang, G.; Yefremenko, V. Measurement of Dielectric Loss in Silicon Nitride at Centimeter and Millimeter Wavelengths. *IEEE Trans. Appl. Supercond.* **2023**, *33*, 1101707.
(86) Skacel, S. T.; Kaiser, C.; Wuensch, S.; Rotzinger, H.; Lukashenko, A.; Jerger, M.; Weiss, G.; Siegel, M.; Ustinov, A. V. Probing the Density of States of Two-Level Tunneling Systems in Silicon Oxide Films Using Superconducting Lumped Element Resonators. *Appl. Phys. Lett.* **2015**, *106*, 022603.
(87) Marcaud, G.; Perello, D.; Chen, C.; Umbarkar, E.; Weiland, C.; Gao, J.; Diez, S.; Ly, V.; Mahuli, N.; D'Souza, N.; He, Y.; Aghaeimeibodi, S.; Resnick, R.; Jaye, C.; Rumaiz, A. K.; Fischer, D. A.; Hunt, M.; Painter, O.; Jarrige, I. Low-Loss Superconducting Resonators Fabricated from Tantalum Films Grown at Room Temperature. *Communications Materials* **2025**, *6*, 182-182.
(88) Poorgholam-Khanjari, S.; Foshat, P.; Zhang, M.; Seferai, V.; Weides, M.; Delfanazari, K. On-Chip Microwave Sensing of Nonequilibrium Quasiparticles in α-Tantalum Superconducting Circuits on Silicon for Scalable Quantum Technologies. *ACS Applied Materials and Interfaces* **2026**, *18*, 3007-3013.
(89) Hammer, G.; Wuensch, S.; Ilin, K.; Siegel, M. Ultra High Quality Factor Resonators for Kinetic Inductance Detectors. *J. Phys.:Conf. Ser.* **2008**, *97*, 012044.

(90) Delfanazari, K. Chip-Scale Electrically Driven Superconducting Coherent Photon Sources for Quantum Information Processing. *Nature Photonics* **2025**, *19*, 1163-1177.
(91) Poorgholam-Khanjari, S.; Seferai, V.; Foshat, P.; Rose, C.; Feng, H.; Hadfield, R. H.; Weides, M.; Delfanazari, K. Engineering High-Q Superconducting Tantalum Microwave Coplanar Waveguide Resonators for Compact Coherent Quantum Circuit. *Sci. Rep.* **2025**, *15*, 27113-27113.
(92) Nelson, A. R. J.; Prescott, S. W. Refnx: Neutron and X-Ray Reflectometry Analysis in Python. *J. Appl. Crystallogr.* **2019**, *52*, 193-200.
(93) Vallières, A.; Russell, M. E.; You, X.; Garcia-Wetten, D. A.; Goronzy, D. P.; Walker, M. J.; Bedzyk, M. J.; Hersam, M. C.; Romanenko, A.; Lu, Y.; Grassellino, A.; Koch, J.; McRae, C. R. H. Loss Tangent Fluctuations Due to Two-Level Systems in Superconducting Microwave Resonators. *Appl. Phys. Lett.* **2025**, *126*, 124001.
(94) Khalil, M. S.; Stoutimore, M. J. A.; Wellstood, F. C.; Osborn, K. D. An Analysis Method for Asymmetric Resonator Transmission Applied to Superconducting Devices. *J. Appl. Phys.* **2012**, *111*, 054510.
(95) Baity, P. G.; Maclean, C.; Seferai, V.; Bronstein, J.; Shu, Y.; Hemakumara, T.; Weides, M. Circle Fit Optimization for Resonator Quality Factor Measurements: Point Redistribution for Maximal Accuracy. *Phys. Rev. Res.* **2024**, *6*, 013329.